\pdfoutput=1%
\documentclass[a4paper,aip,american,citeautoscript,floatfix,jcp,longbibliography,pdftex,superscriptaddress,%
reprint,twocolumn]{revtex4-1}
\usepackage{amsfonts,amsmath,amssymb}%
\usepackage{graphicx}%
\usepackage[utf8]{inputenc}%
\usepackage[T1]{fontenc}
\usepackage{textcomp}%
\usepackage[textsize=footnotesize]{todonotes}%
\usepackage{xspace}
\usepackage[breaklinks=true]{hyperref}%
\usepackage{breakurl}%
\usepackage{hypernat}%
\usepackage{soul}%

\newlength{\figwidth}%
\newlength{\figwidthlarge}%
\newlength{\figwidthsmall}%
\graphicspath{ {./figures/} } 

\newcommand{\mathematicaFig}{0.38}

\usepackage{rotating}
\usepackage{ mathrsfs } 
\usepackage{xspace}
\usepackage{braket}
\usepackage{ upgreek }
\usepackage{siunitx}
\newcommand{\mv}[1]{ \mathbf{#1}}
\newcommand{\ri}[1]{ \mathrm{_{#1}}}
\newcommand{\kd}[2]{\delta_{{#1}{#2}}}

\newcommand{\WignerD}[6]{\mathcal D^{#1} _{{#2}{#3}} (#4,#5,#6)} 

\newcommand{\WignerDop}[3]{\mathcal D^{#1} _{{#2}{#3}} } 

\newcommand{\tj}[6]{ \begin{pmatrix}
  #1 & #2 & #3 \\
  #4 & #5 & #6 
 \end{pmatrix}}
 \newcommand{\sj}[6]{ \begin{Bmatrix}
  #1 & #2 & #3 \\
  #4 & #5 & #6 
 \end{Bmatrix}}
 \newcommand{\ninej}[9]{ \begin{Bmatrix}
  #1 & #2 & #3 \\
  #4 & #5 & #6 \\
  #7 & #8 & #9
 \end{Bmatrix}}
\newcommand{\ntwopXstate}{X$^2\Sigma_\mathrm{g}^+$\xspace}

\newcommand{\ntp}{N$_{2}^{+}$\xspace}

\newcommand{\np}{n^+}
\newcommand{\vp}{v^+}
\newcommand{\Jp}{J^+}
\newcommand{\Fp}{F^+}
\newcommand{\MJp}{M^+_J}
\newcommand{\MJ}{M_J}
\newcommand{\MF}{M_F}
\newcommand{\MFp}{M_F^+}
\newcommand{\MNp}{M^+_N}
\newcommand{\MSp}{M^+_S}
\newcommand{\Np}{N^+}
\newcommand{\Sp}{S^+}
\newcommand{\Lp}{\Lambda^+}
\newcommand{\MN}{M_N}
\newcommand{\MS}{M_S}
\newcommand{\Ip}{I^+}
\newcommand{\MI}{M_I}
\newcommand{\MIp}{M_I^+}
\newcommand{\CG}[6]{C_{#1 #2 #3 #4} ^{#5 #6}}
\newcommand{\bBetaJ}{(b$_{\beta_J}$)\xspace}
\newcommand{\OtwoXstate}{X$^3\Sigma^{-}_\mathrm{g}$ \xspace}
\newcommand{\OtwoPlusbstate}{b$^4\Sigma^{-}_\mathrm{g}$\xspace}
\newcommand{\OtwoPlus}{O$_2^+$\xspace}
\newcommand{\upr}{w}
\newcommand{\formspht}[2]{\mathrm{T}^{#1}_{#2}}
\newcommand{\formsphtMolFixed}[2]{\mathrm{T'}^{#1}_{#2}}

\newcommand{\myhat}[1]{#1}

\begin{document}
\title{Fine- and hyperfine-structure effects in molecular photoionization: I. General theory and direct photoionization}%
\author{Matthias Germann}
\author{Stefan Willitsch}\email{stefan.willitsch@unibas.ch}
\affiliation{Department of Chemistry, University of Basel, Klingelbergstrasse 80, 4056 Basel, Switzerland}
\date{\today}%
\begin{abstract}\noindent%
We develop a model for predicting fine- and hyperfine intensities in the direct photoionization of molecules based on the separability of electron and nuclear spin states from vibrational-electronic states. Using spherical tensor algebra, we derive highly symmetrized forms of the squared photoionization dipole matrix elements from which which we derive the salient selection and propensity rules for fine- and hyperfine resolved photoionizing transitions. Our theoretical results are validated by the analysis of the fine-structure resolved photoelectron spectrum of O$_2$ (reported by H. Palm and F. Merkt, Phys. Rev. Lett. {\bf 81}, 1385 (1998)) and are used for predicting hyperfine populations of molecular ions produced by photoionization. 
\end{abstract}
\maketitle%


\section{Introduction}
\label{sec:intro}

Photoionization and photoelectron spectroscopy are among the eminent experimental techniques to gain information on the electronic structure of molecules, on their photoionization dynamics and the structure and dynamics of molecular cations \cite{turner70a, ellis05a, merkt11a}. Since the first photoelectron spectroscopic studies of molecules in the 1960s, the resolution of the technique has steadily been improved, such that vibrational and thereon rotational structure were resolved during the following decades \cite{willitsch05b, merkt11a}. Along with the experimental progress, the theoretical understanding of molecular photoionization has been successively refined. Vibrational structure in the spectra can often be modeled in terms of the Franck-Condon principle (see, e.g., Ref. \onlinecite{ellis05a}). Concerning rotational structure, Buckingham, Orr and Sichel (BOS) presented in a seminal paper \cite{buckingham70a} in 1970 a model to describe rotational line intensities in photoelectron spectra of diatomic molecules. Subsequently, extended models, e.g., for resonance-enhanced multiphoton ionization \cite{dixit85a} or to describe the angular distribution of the photoelectrons \cite{allendorf89a,wang91a}, have been developed. Also, the BOS model has been rephrased in terms of spherical tensor algebra \cite{xie92a} and extended to asymmetric rotors \cite{willitsch05b}. Moreover, dipole selection rules for photoionizing transitions have been developed based on these intensity models \cite{dixit86a} as well as on general symmetry considerations \cite{xie90a, signorell97c}.

Over the last decades, fine (spin-rotational) structure has been resolved in high-resolution photoelectron spectra \cite{palm98c}. Hyperfine structure has been resolved in millimeter-wave spectra of high Rydberg states of rare gas atoms such as Kr and diatomics such as H$_2$ and its isotopomers, see, e.g., Refs. \onlinecite{woerner03a,osterwalder04a,cruse08a,jungen11a,sprecher14a}. Fine- and hyperfine-structure effects in molecular photoionization have also become of importance in precision spectroscopy and dynamics experiments with molecular ions produced by photoionization in which the ionic hyperfine populations are governed by the underlying hyperfine photoionization dynamics \cite{tong10a,tong12a, germann14a}. Thus, there is a growing need for theoretical models capable of describing fine- and hyperfine effects in molecular photoionization. Whereas the hyperfine structure in Rydberg spectra has previously been treated within the framework of multichannel quantum-defect theory (MQDT) \cite{woerner03a,osterwalder04a,cruse08a,jungen11a,sprecher14a}, we are not aware of any previous treatments of hyperfine intensities in direct photoionization which can be applied to the interpretation of line intensities in high-resolution photoelectron spectra and to the prediction of level populations of molecular cations produced by photoionization. The present work aims at filling this gap by developing closed expressions for fine- and hyperfine-structure resolved intensities in molecular photoionization. The theory is developed here for diatomic molecules in Hund's coupling case (b), but can readily be extended to other coupling cases by a suitable basis transformation\cite{brown03a} of our final result or to symmetric- and asymmetric-top molecules by a suitable modification of the rotational basis functions \cite{willitsch05b}. 

In the present paper, we develop the general theory for fine- and hyperfine-resolved photoionization intensities and apply our model to the analysis of the fine structure of the photoelectron spectrum of O$_2$ from Ref.~\onlinecite{palm98c} and of hyperfine propensities in the photoionization of N$_2$. In the subsequent companion paper\cite{germann16cPROVISORISCH_arXiv}, we extend our model to resonance-enhanced multiphoton-ionization processes and address the problem of hyperfine-preparation of molecular cations.


\section{General considerations} 
\label{sec:general}

We consider the ionization of a molecule M yielding the molecular ion M$^{+}$ by ejection of a photoelectron e$^{-}$ through interaction with electromagnetic radiation via the electric-dipole operator $\boldsymbol{\myhat\mu} $:
\begin{equation}
\text{M} \; \xrightarrow{\parbox{0.75cm}{\centering $\boldsymbol{\myhat\mu}$  }  }  \; \text{M}^{+} + \text{e}^{-}.
\end{equation}

The relevant transition matrix element is given by
\begin{equation}
\left ( \bra{ \psi_{\mathrm{M}^{+}}} \bra{\psi_{\mathrm e^{-}}} \right ) \boldsymbol{\myhat\mu}   \ket{\psi_\mathrm{M}} ,
\end{equation}
with $\ket{ \psi_\mathrm{M} }$ standing for the internal quantum state of the neutral molecule and $\ket{ \psi_{\mathrm{M}^{+}} }$$\ket{\psi_{\mathrm e^{-}}} $ the product of the internal state of the molecular ion M$^{+}$ and the state of the photoelectron $\mathrm{e}^-$.

The squared magnitude of the transition matrix element summed over all the quantum states contributing to the observed ionization rate,
\begin{equation}
P(\mathrm{M} \rightarrow\mathrm{M}^{+})
=
\sum_{\psi_{\mathrm{M}}} \sum_{\psi_{\mathrm{M}^{+}}} \sum_{ \psi_{\mathrm e^{-}} }
\big |
\left (
\bra{ \psi_{\mathrm{M}^{+}}} \bra{\psi_{\mathrm e^{-}}} 
\right )
  \boldsymbol{\myhat\mu} 
    \ket{\psi_\mathrm{M}} 
\big |^{2}, \label{eq:piprob}
\end{equation}
is proportional  to the ionization probability per unit time. Here, the sums over $\psi_{\mathrm{M}}$ and $\psi_{\mathrm{M}^{+}}$ include all degenerate (or spectroscopically unresolved) states of the neutral molecule and the molecular ion, respectively, involved in the photoionizing transition. The sum over $\psi_{\mathrm e^{-}}$ includes the orbital angular momentum and the spin state of the emitted photoelectron. We suppose that neither the energy nor the angular distribution nor the spin polarization of the photoelectron is detected in the photoionization experiment. 

\begin{table*}[tpb]
\centering
\caption{
Angular momentum quantum numbers relevant to the photoionization of diatomic molecules}
\label{tab:QNs_direct_ionization}
\begin{footnotesize}
\newlength{\qnTableSpace}
\setlength{\qnTableSpace}{0.7mm}
\begin{tabular}{llll}
\shortstack[l]{Magnitude \\ quant. num.}  & \shortstack[l]{Mol.-fixed\\projection} & \shortstack[l]{Space-fixed\\projection} & Description                                                                                                       \\[\qnTableSpace] 
\hline
\rule{0mm}{3.5mm}$ N $            & $ \Lambda $               & $ \MN $                & Orbital-rotational angular momentum of the neutral molecule                                                        \\[\qnTableSpace]
$ S $            & --                         & $ \MS $                & Total electron spin of the neutral molecule                                                                        \\[\qnTableSpace]
$ J $            & --                         & $ \MJ $                & Total angular momentum of the neutral molecule excluding nuclear \\
&&									&  spin \\[\qnTableSpace]
$ I $            & --                         & $ \MI $                & Nuclear spin of the neutral molecule                                                                              \\[\qnTableSpace]
$ F $            & --                         & $ \MF $                & Total angular momentum of the neutral molecule                         \\[\qnTableSpace]
$ \Np $          & $ \Lp $                   & $ \MNp $               & Orbital-rotational angular momentum of the molecular ion                                                       \\[\qnTableSpace]
$ \Sp $          & --                         & $ \MSp $               & Total electron spin of the molecular ion                                                                           \\[\qnTableSpace]
$ \Jp $          & --                         & $ \MJp $               & Total angular momentum of the molecular ion  excluding nuclear      \\
&&									&   spin \\[\qnTableSpace]
$ \Ip $           & --                         & $ \MIp $               & Nuclear spin of the molecular ion                                                                                  \\[\qnTableSpace]
$ \Fp $            & --                         & $ \MFp $                & Total angular momentum of the molecular ion                            \\[\qnTableSpace]
$ l $            & --                         & $ m_l $                & Orbital angular momentum (partial wave) of the photoelectron                                                                      \\[\qnTableSpace]
$ s $            & --                         & $ m_s $                & Spin of the photoelectron ($s=1/2$)                                                                                         \\[\qnTableSpace]
1                & --                         & $\mu_0$                & Angular momentum due to the electric-dipolar interaction                            \\
&& 									& with the electromagnetic field \\[\qnTableSpace]
$ k $            & $ q $                     & $ p $                  & Total orbital angular momentum transferred to/from the molecule         \\
&& 									& in the ionization process ($ p=-m_l+\mu_0 $)  \\[\qnTableSpace]
$ u $            & --                        & $\upr$                  & Total angular momentum transferred to/from         \\
&&									&  the molecule in the ionization process ($ \upr = -m_s + p $) \\[\qnTableSpace]        
\hline
\end{tabular}
\end{footnotesize}
\end{table*}

\section{Fine-structure resolved photoionization intensities}
\label{sec:FS_effects_direct_ionization}

We start by developing the theory for fine-structure-resolved photoionization intensities which will form the basis for the subsequent inclusion of hyperfine structure in Sec.~\ref{sec:hfs}. We note that a similar result for fine-structure resolved photoionization has previously been derived by McKoy and co-workers \cite{dixit85a, braunstein92a}. In Hund's case (b), fine structure manifests itself in the coupling of the orbital-rotational angular momentum $\myhat{\mv{N}}$ (corresponding to the mechanical rotation of the molecule in $\Sigma$ states) with the electron spin $\myhat{\mv{S}}$ to form $\myhat{\mv{J}}$:  $\myhat{\mv{J}} = \myhat{\mv{N}} + \myhat{\mv{S}}$.

Therefore, we evaluate Eq.~\eqref{eq:piprob} for a diatomic molecule in Hund's case (b) by expressing the quantum states of the neutral molecule $\ket{ \psi_\mathrm{M} }$ and the molecular ion $\ket{ \psi_{\mathrm{M}^{+}} }$ in the basis $ \Ket{n \Lambda, v, N \Lambda S J \MJ}$ and $\Ket{\np \Lp, \vp, \Np \Lp \Sp \Jp \MJp} $, respectively. The definitions of the relevant angular-momentum quantum numbers are summarized in Tab.~\ref{tab:QNs_direct_ionization}. $n$ ($\np$) and $v$ ($\vp$) denote the electronic and the vibrational quantum number in the neutral molecule (molecular ion). The state of the photoelectron $\ket{\psi_{\mathrm e^{-}}} $ is expressed as a tensor product of its spin state and its partial wave $ \Ket{s,m_s}  \Ket{l,m_l}$. 

The quantity $P(J,\Jp)$ which is proportional to the ionization probability on the photoionizing transition $J\rightarrow J^+$ may hence be written as
\begin{widetext}
\begin{align}
P(J,\Jp) 
=
\sum _{l=0}^{\infty} \sum _{m_l=-l}^{l}  \sum _{m_s=-s}^{s} \sum_{M_J=-J}^{J}\sum_{\MJp=-\Jp}^{\Jp}  
  \Big | 
  \left (
  \Bra{\np \Lp, \vp, \Np \Lp \Sp \Jp \MJp}  \Bra{s,m_s}  \Bra{l,m_l} 
 \right )
 \boldsymbol{\myhat\mu}
   \Ket{n \Lambda, v, N \Lambda S J \MJ} \Big  |^2 .
 \label{eq_fs_transition_probability}
\end{align}
\end{widetext}
We follow the approach of Xie and Zare \cite{xie92a} and identify the electric-dipole operator $\boldsymbol{\myhat\mu}$ and the photoelectron partial wave $\Ket{l,m_{l}}$ with the spherical tensors $\formspht{1}{\mu_{0}}$ and $\formspht{l}{-m_{l}} $, respectively. We then contract the product of these two spherical tensors \cite{zare88a,varshalovich89a,brown03a} according to
\begin{equation}
\formspht{l}{-m_{l}} \otimes \formspht{1}{\mu_{0}} =  \sum _{k=| l-1|} ^{l+1} \CG{l}{-m_l}{1}{\mu_0}{k}{p} \formspht{k}{p},
\label{eq_mu_and_l_coupled}
\end{equation}
where $p = - m_l + \mu_{0}$ and $\mu_0$ denotes the polarization state of the photon. $\CG{j_1}{m_1}{ j_2}{m_2}{j}{m}$ stands for a Clebsch-Gordan coefficient \cite{edmonds64a,zare88a,varshalovich89a}. The spherical tensor operator $\formspht{k}{p}$ of Eq.~\eqref{eq_mu_and_l_coupled} describes the combined effect of absorbing electromagnetic radiation via the electric dipole operator and ejecting a photoelectron in the state $\Ket{l,m_l}$ (with $l=k\pm1$).

The term with $k=l$ does not contribute to the sum in Eq.~\eqref{eq_fs_transition_probability} because of parity selection rules and may be omitted \cite{xie92a,xie90a}. Ignoring proportionality constants, the matrix element in Eq.~\eqref{eq_fs_transition_probability} is thus expressed as
\begin{widetext}
\begin{multline}
\left (
\Bra{\np \Lp, \vp, \Np \Lp \Sp \Jp \MJp}  \Bra{s,m_s}  \Bra{l,m_l}
\right )
\boldsymbol{\myhat\mu}
 \Ket{n \Lambda, v, N \Lambda S J \MJ}
\\ =   \sum _{k=l\pm1}  \CG{l}{-m_l}{1}{\mu_0}{k}{p}
\left (
\Bra{\np \Lp, \vp, \Np \Lp \Sp \Jp \MJp}  \Bra{s,m_s} 
\right )
\formspht{k}{p}
\Ket{n \Lambda, v, N \Lambda S J \MJ},
\label{eq_fs_transition_mat_el}
\end{multline}
\end{widetext}
where negative values for $k$ are to be excluded.

To proceed, we decouple spin and orbital-rotational angular momenta in the neutral state according to
\begin{multline}
\Ket{n \Lambda,v,N \Lambda S J \MJ}  \\
=  \sum_{\MN,\MS} \CG{N}{\MN}{S}{\MS}{J}{\MJ} \Ket{n \Lambda,v, N \Lambda \MN, S \MS},
 \end{multline}
and analogously in the ionic state. The transition matrix element in Eq.~\eqref{eq_fs_transition_mat_el} then reads, 
\begin{align}
& \left ( \Bra{\np \Lp,  \vp, \Np \Lp \Sp \Jp \MJp}  \Bra{s,m_s} \right )
 \nonumber \\ & \; \; \; \; \; \; \; \; \formspht{k}{p}  \Ket{n \Lambda, v, N \Lambda S J \MJ}
\nonumber \\ & \; \; \; \; \; \; \; \;
 \begin{aligned}  = & \sum_{\MN,\MS} \sum_{\MNp,\MSp} \CG{N}{\MN}{S}{\MS}{J}{\MJ}  \CG{\Np}{\MNp}{\Sp}{\MSp}{\Jp}{\MJp}
\\
& \times \left (  \Bra{\np \Lp,\vp, \Np \Lp \MNp, \Sp \MSp }  \Bra{s,m_s} \right )
\\
& \times \formspht{k}{p} \Ket{n \Lambda,v, N \Lambda \MN, S \MS}.
\end{aligned}
\end{align}

Since the tensor operator $\formspht{k}{p}$ does not operate on the spin functions, the matrix element on the last and next-to-last line above can be separated into a rotational-vibronic and a pure spin factor:
\begin{multline}
\left (
\Bra{\np \Lp,\vp, \Np \Lp \MNp, \Sp \MSp }  \Bra{s,m_s}
\right )
 \\ 
\times  \formspht{k}{p} \Ket{n \Lambda,v, N \Lambda \MN, S \MS} = \\ 
\Braket{\np \Lp,\vp, \Np \Lp \MNp | \formspht{k}{p} | n \Lambda,v, N \Lambda \MN}  \\
\times  \left (\Bra{\Sp \MSp} \Bra{s,m_s} \right) \Ket{S \MS}.
\label{eq_fs_spin_rot_separated}
\end{multline}

In order to calculate the rotational-vibronic factor on the second last line in Eq.~\eqref{eq_fs_spin_rot_separated}, we transform the spherical tensor operator $\formspht{k}{p}$ from space-fixed to molecule-fixed coordinates using Wigner rotation matrix elements
\begin{equation}
\formspht{k}{p} = \sum_{q=-k}^k [\WignerDop{k}{p}{q}]^{*} \formsphtMolFixed{k}{q},
\end{equation}
where the $^{\prime}$ denotes operators in molecule-fixed coordinates.

The Wigner rotation matrix elements only act on the angular coordinates, whereas the tensor $\formsphtMolFixed{k}{q}$ only operates on the vibronic state. Therefore, we can write:
\begin{align}
&\Braket{\np \Lp,\vp, \Np \Lp \MNp | \formspht{k}{p} | n \Lambda,v, N \Lambda \MN}
\nonumber \\ & \; \; \; \; \; \; \; \;  =  \sum_{q=-k}^k \Braket{\np \Lp,\vp | \formsphtMolFixed{k}{q} | n \Lambda,v}
\nonumber \\ &  \qquad\qquad\qquad \times  \Braket{\Np \Lp \MNp |  [\WignerDop{k}{p}{q}]^{*} | N \Lambda \MN}.
\label{eq_fs_vibronic_rot_separated}
\end{align}

Upon substituting the rotational states of the neutral molecule by Wigner rotation matrices
\begin{equation}
\Braket{\phi \, \theta \, \chi | N \Lambda \MN} = \sqrt{\frac{2N+1}{8\pi^2}}  \left[ \WignerD{(N)}{\MN}{\Lambda}{ \phi}{\theta}{\chi} \right]^{*},
\end{equation}
and analogously for the ion, we obtain for the last line in Eq.~\eqref{eq_fs_vibronic_rot_separated} an integral over a product of three Wigner rotation matrices over the Euler angles $\phi,\theta,\chi$ which accounts for \cite{edmonds64a,zare88a,brown03a}:
\begin{multline}
\Braket{\Np \Lp \MNp |  [\WignerDop{k}{p}{q}]^{*} | N \Lambda \MN}
\\ = \sqrt{2 \Np +1} \sqrt{2N+1} (-1)^{\MNp-\Lp} \\ 
\times \tj{\Np}{k}{N}{-\MNp}{p}{\MN} \tj{\Np}{k}{N}{-\Lp}{q}{\Lambda}.
\end{multline}
Because of the second Wigner 3j-symbol, this expression vanishes for all values of $q$ but $q= {\Lp - \Lambda =: \Delta \Lambda}$. Hence, only this value contributes to the sum in Eq.~\eqref{eq_fs_vibronic_rot_separated} and we may write 
\begin{align}
&\Braket{\np \Lp,\vp, \Np \Lp \MNp | \formspht{k}{p} | n \Lambda,v, N \Lambda \MN} \nonumber \\ 
&= \Braket{\np \Lp,\vp | \formsphtMolFixed{k}{\Delta\Lambda} | n \Lambda,v} \sqrt{2 \Np +1} \sqrt{2N+1} \nonumber \\
 & \ \ \ \times (-1)^{\MNp-\Lp} \tj{\Np}{k}{N}{-\MNp}{p}{\MN} \tj{\Np}{k}{N}{-\Lp}{\Delta\Lambda}{\Lambda}.
\end{align}

To compute the spin part of Eq.~\eqref{eq_fs_spin_rot_separated}, we couple the spin of the ion and the photoelectron to get the total electronic spin after ionization:
\begin{multline}
\Bra{\Sp \MSp, s m_s} \\
= \sum_{S\ri{tot} = | \Sp - s|}^{\Sp + s} \sum_{M_{S\ri{tot}} = - S\ri{tot}}^{S\ri{tot}} 
\CG{\Sp}{\MSp}{s}{m_s}{S\ri{tot}}{M_{S\ri{tot}}} \Bra{S\ri{tot} M_{S\ri{tot}}}.
\end{multline}
Assuming orthonormal spin states, we thus obtain for the spin factor in Eq.~\eqref{eq_fs_spin_rot_separated}:
\begin{align}
& \Braket{\Sp \MSp, s m_s | S \MS} \nonumber \\
& = \sum_{S\ri{tot} = | \Sp - s|}^{\Sp + s}  \sum_{M_{S\ri{tot}} = - S\ri{tot}}^{S\ri{tot}} \CG{\Sp}{\MSp}{s}{m_s}{S\ri{tot}}{M_{S\ri{tot}}} \Braket{S\ri{tot} M_{S\ri{tot}} | S \MS} \\ 
& = \CG{\Sp}{\MSp}{s}{m_s}{S}{\MS}.
\end{align}

Collecting these results and substituting them into Eq.~\eqref{eq_fs_transition_mat_el}, we obtain the matrix element for spin-rotation-resolved photoionization dipole transitions as:
\begin{widetext}
\begin{align}
&
\left (
\Bra{\np\Lp,  \vp, \Np \Lp \Sp \Jp \MJp}  \Bra{s,m_s}  \Bra{l,m_l} 
\right )
\boldsymbol{\myhat\mu}  \Ket{n \Lambda, v, N \Lambda S J \MJ}
\nonumber \\ & \begin{aligned} 
\; \; \; \; \; \; \; \;   = \, &
\sqrt{2 \Np +1} \sqrt{2N+1} \sqrt{2S+1} \sqrt{2\Jp+1} \sqrt{2J+1} (-1)^{l-1+p+N+\Np-\Lp-S-s+\MJp+\MJ}
\\& \times \sum _{k=l\pm 1} \sqrt{2k+1}  \tj{l}{ 1}{k}{ -m_l}{ \mu_0}{ -p}
\Braket{\np \Lp, \vp | \formsphtMolFixed{k}{\Delta\Lambda} | n \Lambda,v}
\tj{\Np}{k}{N}{-\Lp}{\Delta\Lambda}{\Lambda}
\\ & \times \sum_{\MSp,\MS} \sum_{\MNp,\MN} (-1)^{\MNp+\MS} \tj{\Np}{\Sp}{\Jp}{\MNp}{\MSp}{-\MJp} \tj{N}{S}{J}{\MN}{\MS}{-\MJ}
\\ & \times \tj{\Np}{k}{N}{-\MNp}{p}{\MN}  \tj{\Sp}{s}{S}{\MSp}{m_s}{-\MS} ,
\end{aligned}
\label{eq_fs_complete_trans_mat_el_not_simplified}
\end{align}
\end{widetext}
where the Clebsch-Gordan coefficients have been replaced by Wigner 3j-symbols. 

In principle, Eq.~\eqref{eq_fs_complete_trans_mat_el_not_simplified} completely describes the matrix element for fine-structure-resolved photoionization transitions and could---when substituted into Eq.~\eqref{eq_fs_transition_probability}---be used for analyzing measured photoionization and photoelectron spectra and predicting relative photoionization intensities. However, the complexity of this expression complicates a deeper insight into the physics of the photoionization process and the multiple sums render the evaluation of this expression computationally expensive.

However, using the properties of the Wigner 3j-symbols, the terms on the last and next-to-last line in Eq.~\eqref{eq_fs_complete_trans_mat_el_not_simplified}, may be expressed in from of a Wigner 9j-symbol\cite{edmonds64a,zare88a,varshalovich89a} (the expression in curly brackets below). In this way, the sums over $\MN$, $\MNp$, $\MS$ and $\MSp$ are avoided and the matrix element becomes\cite{germann16a},
\begin{align}
&
\left (
\Bra{\np\Lp, \vp, \Np \Lp \Sp \Jp \MJp}  \Bra{s,m_s}  \Bra{l,m_l}
\right )
\nonumber \\
& \myhat{\boldsymbol\mu}  \Ket{n \Lambda, v, N \Lambda S J \MJ}
\nonumber \\ & 
\begin{aligned}
 \; \; \;   = \, &
\sqrt{2 \Np +1} \sqrt{2N+1}  \sqrt{2 S +1}  \sqrt{2 \Jp +1} \sqrt{2J+1}
\\  & \times
(-1)^{l -1 - \Lp + N + J - s + 2\MJ + \MJp} \sum _{k=l\pm 1} (-1)^{k} \sqrt{2k+1}
\\ & \times
 \tj{l}{ 1}{k}{ -m_l}{ \mu_0}{ -p} \Braket{\np \Lp, \vp | \formsphtMolFixed{k}{\Delta\Lambda} | n \Lambda, v} 
\\ & \times
\tj{\Np}{k}{N}{-\Lp}{\Delta\Lambda}{\Lambda} 
\sum_{u=|k-s|}^{k+s}(2u+1) \tj{\Jp}{u}{J}{-\MJp}{\upr}{\MJ}
\\ & \times
\tj{u}{k}{s}{\upr}{-p}{m_s}  \ninej{\Jp}{u}{J}{\Np}{k}{N}{\Sp}{s}{S}.
\end{aligned}
\label{eq_fs_complete_trans_mat_el_with_9j}
\end{align}
Here, the angular momentum quantum number $u$ with the associated space-fixed projection $\upr$ (given by $\upr = - m_s + p$) has been introduced. $u$ represents the resultant of the coupling of $k$ and $s$. Its physical meaning is described further below.

Substituting this matrix element into Eq.~\eqref{eq_fs_transition_probability} and simplifying the result, we obtain the quantity $P(J,\Jp)$ as:
\begin{align}
P(J,\Jp) = \: &(2 \Np +1) (2N+1) (2 S +1) (2 \Jp +1) (2J+1) \nonumber \\
 &\times \sum _l  \sum _{k=l\pm 1} (2k+1)  \tj{\Np}{k}{N}{-\Lp}{\Delta\Lambda}{\Lambda}^2 \nonumber \\
 & \times \left | \Braket{\np \Lp, \vp | \formsphtMolFixed{k}{\Delta\Lambda} | n \Lambda, v}\right |^2 \nonumber \\
  &\times \sum_{u=|k-s|}^{k+s}(2u+1)   \ninej{\Jp}{u}{J}{\Np}{k}{N}{\Sp}{s}{S}^2 \nonumber \\
& \times  \sum _{m_l}\tj{l}{ 1}{k}{ -m_l}{ \mu_0}{ -p}^2  \sum _{m_s} \tj{u}{k}{s}{\upr}{-p}{m_s}^2.
\label{eq_fs_transition_probability_no_cross_terms}
\end{align}
Owing to the orthogonality properties of the Wigner 3j-symbols\cite{edmonds64a,zare88a,varshalovich89a}, the cross terms in the above expression vanish when summed over all possible values for $\MJ$ and $\MJp$ resulting in a particularly simple form for Eq.~\eqref{eq_fs_transition_probability_no_cross_terms}.

For linear polarized radiation, as is used in most experiments, we have $\mu_{0} = 0$ in a suitably chosen coordinate system. The sums over $m_{l}$ and $m_{s}$ then account for $1/(3(2k+1))$ and we get as a final result:
\begin{align}
P(J,\Jp) = \: & \frac{1}{3}(2 \Np +1) (2N+1) (2 S +1) (2 \Jp +1)  \nonumber \\
& \times (2J+1) \sum _l  \sum _{k=l\pm 1} \tj{\Np}{k}{N}{-\Lp}{\Delta\Lambda}{\Lambda}^2 \nonumber \\ 
& \times \left | \Braket{\np \Lp, \vp | \formsphtMolFixed{k}{\Delta\Lambda} | n \Lambda, v}\right |^2 \nonumber \\
& \times \sum_{u=|k-s|}^{k+s}(2u+1)   \ninej{\Jp}{u}{J}{\Np}{k}{N}{\Sp}{s}{S}^2.
\label{eq_fs_transition_probability_lin_pol}
\end{align}

The highly symmetrized form of  Eq.~\eqref{eq_fs_transition_probability_lin_pol} allows a detailed and insightful physical interpretation and discussion of photoionization selection rules. The photoelectron ejected in the ionization process is described by a partial wave $l$. The probability of a transition from a specific neutral to an ionic state is obtained by summing over all partial waves. The photoelectron partial waves allowed for a particular electronic state of the ion and the neutral precursor molecule are constrained by the parity of these states \cite{signorell97c, xie90a, xie92a}. If neutral and ionic states have the same parity ($\pm \leftrightarrow \pm$ transitions), only odd values of $l$ occur: $l=1,3,5,\dots$ In the case of unequal parities of neutral and ionic state ($\pm \leftrightarrow \mp$), only even $l$ values are allowed: $l=0,2,4,\dots$

In addition to the angular momentum carried by the departing photoelectron, the molecule also exchanges angular momentum with the electromagnetic wave as described by the electric-dipole operator. The angular momenta associated with the photoelectron partial wave and the dipole excitation are connected with $k$, see Eq.~\eqref{eq_mu_and_l_coupled}. Since the dipole operator is a spherical tensor of first rank, $k$ is constrained to the values $k=l-1$ or $k=l+1$ with $k=l$ forbidden because of parity selection rules \cite{xie92a,xie90a}. We thus have $k=0,2,4,\dots$ for $\pm \leftrightarrow\pm$ transitions and $k=1,3,5,\dots$ for $\pm \leftrightarrow \mp$ transitions \cite{xie92a}.

The interplay between the different angular momenta is expressed by the 9j-symbol in Eq.~\eqref{eq_fs_transition_probability_lin_pol} in a compact way. All rows and columns have to fulfill the triangle inequality for the coupling of angular momenta \cite{zare88a,varshalovich89a}. The couplings in the ionic and the neutral state are expressed by the first and the last column, respectively. The coupling of the angular momenta transferred to or from the molecule is described by the middle column: the spin of the photoelectron $s$ is coupled to $k$ (the angular momentum associated with the photoelectron and the interaction with the electromagnetic field) to form $u$. The rows of the 9j-symbol represent the angular momentum transfer during photoionization: according to the first row, $u$ determines the change in the total angular momentum (excluding nuclear spin),
whereas $k$ decides on the change in the orbital-rotational angular momentum (second row). Finally, $s$ accounts for the electron-spin change in the molecule resulting from the ionization process, as seen from the last row in the 9j-symbol.

The value of $k$ determines the maximum change of the orbital-rotational angular momentum in the photoionization process, i.e., $|\Delta N| =| \Np-N| \le k$, as may be seen from the 3j-symbol or the middle row of the 9j-symbol in Eq.~\eqref{eq_fs_transition_probability_lin_pol}. Transitions with $k=0$ do not allow any change in the orbital-rotational angular momentum, i.e., $\Np=N$. For transitions with $k=2$, the values $\Delta N = 0, \pm 1$ and $\pm 2$ are possible (with $\Delta N = \pm 1$ forbidden for $\Sigma$-$\Sigma$-transitions). The permissible values of $k$ (and therefore the range of photoelectron partial waves $l$) and their weighting factors are determined by the vibronic coefficients $ C_k:=  \left | \Braket{\np \Lp, \vp | \formsphtMolFixed{k}{\Delta\Lambda} | n\Lambda, v} \right | ^2$. To calculate these coefficients, the electronic structure of the molecular ion and its neutral precursor must be known. Thus, the coefficients $C_k$ may either be obtained from an ab-initio calculation or by fitting Eq.~\eqref{eq_fs_transition_probability_lin_pol} to measured photoionization intensities \cite{xie92a,willitsch05b}. In many cases, only one or a few $C_k$ contribute substantially to the total transition probability. If so, only a few free parameters are needed to describe the intensities in a rotationally resolved photoelectron spectrum. 

Within the orbital ionization model \cite{buckingham70a, willitsch05b} which assumes that the photoelectron is ejected from a single molecular orbital and that the orbital structure does not change upon ionization, $k$ assumes the intuitive physical interpretation as the quantum number of the electronic orbital angular momentum left behind in the molecule after photoionization. In this approximation, the permissible values of $k$ and the vibronic coefficients $C_k$ can be obtained from a single-center expansion of the molecular orbital from which the photoelectron is ejected.

Moreover, the photoelectron spin $s =1/2$ is coupled to $k$ to form $u$. The possible values for $u$ are thus $u=|k-1/2|$ and $u=k+1/2$. Similar to $k$ determining the change in the orbital-rotational angular momentum, 
$u$ determines the change in the total angular momentum (excluding nuclear spin) according to the selection rule $|\Delta J| = |\Jp-J| \le u$, as inferred from the first row of the 9j-symbol in Eq.~\eqref{eq_fs_transition_probability_lin_pol}. For, e.g., $k=0$ only $u=1/2$ is possible and thus only transitions with $| \Delta J | = 1/2$ are allowed. For $k=2$, on the other hand, $u=3/2$ and $u=5/2$ are possible, allowing values of $| \Delta J |$ up to 5/2.

Since the values of $u$ are determined by the values of $k$, the vibronic coefficients $C_k$ describing the relative intensities of different rotational lines in a photoelectron spectrum also determine the relative intensities of transitions connecting different fine structure components. Hence, the present model describes the spin-rotational effects in photoionization without additional free parameters.

Eq.~\eqref{eq_fs_transition_probability_lin_pol} also provides a substantially more efficient way for numerical calculations of the transition probabilities as compared to direct evaluation of the transition matrix element in Eq.~\eqref{eq_fs_complete_trans_mat_el_not_simplified} and summation over angular momentum projection quantum numbers (as in Eq.~\eqref{eq_fs_transition_probability}), since the number of computationally demanding multiple sums is minimized.

Eq.~\eqref{eq_fs_transition_probability_lin_pol} may be compared with the similar result given in Eq.~(6) of Dixit et al.\@ \cite{dixit85b}. The equivalence of the two results can be seen by taking the squared absolute magnitude of Eq.~(6) in Ref. \onlinecite{dixit85b}, integrating over the entire unit sphere and making use of orthogonality properties of Wigner symbols and spherical harmonics.

We note that Eq.~\eqref{eq_fs_transition_probability_lin_pol} can also be directly applied to symmetric top-molecules exhibiting a Hund's case (b)-type coupling hierarchy by substituting the quantum numbers $\Lambda, \Lambda^+$ by the symmetric-top quantum numbers $K, K^+$. Asymmetric top molecules can be treated by a suitable substitution of the rotational wavefunction as shown in Ref. \onlinecite{willitsch05b}. Moreover, other Hund's cases can be treated by an appropriate frame transformation of Eqs.~\eqref{eq_fs_transition_probability_no_cross_terms} and \eqref{eq_fs_transition_probability_lin_pol}, see, e.g., Ref. \onlinecite{brown03a}. 

Often, the neutral molecule is in a singlet state, i.e, $S=0$. In this case, we have $J=N$ and $\Sp=s=1/2$ and our result may be simplified further. The 9j-symbol then equals a 6j-symbol \cite{zare88a,varshalovich89a},
\begin{equation}
 \ninej{\Jp}{u}{N}{\Np}{k}{N}{\Sp}{s}{0}^2
  =
  \frac{1}{2(2N+1)} \sj{\Jp}{u}{N}{k}{\Np}{s}^{2},
\end{equation}
and Eq.~\eqref{eq_fs_transition_probability_lin_pol} becomes,
\begin{align}
P_{S=0}(N,\Jp) &= \frac{1}{6} (2N+1) (2 \Np +1) (2 \Jp +1)  \nonumber \\
&\times  \sum _l  \sum _{k=l\pm 1} \tj{\Np}{k}{N}{-\Lp}{\Delta\Lambda}{\Lambda}^2  \nonumber \\
&\times \left | \Braket{\np \Lp, \vp | \formsphtMolFixed{k}{\Delta\Lambda} | n \Lambda, v}\right |^2 \nonumber \\
&\times  \sum_{u=|k-s|}^{k+s}(2u+1)    \sj{\Jp}{u}{N}{k}{\Np}{s}^2.
\label{eq_fs_transition_probability_lin_pol_singlet}
\end{align}


\section{Hyperfine-structure resolved photoionization intensities}
\label{sec:hfs}

The above treatment is now extended to photoionizing transitions connecting hyperfine levels. To that end, we need to consider the role of the nuclear spin in the neutral and ionic levels. We assume that both of these levels may be described with the Hund's case \bBetaJ angular momentum coupling scheme \cite{frosch52a,dunn72a,brown03a}. In this scheme, the total nuclear spin $\myhat{\mv{I}}$ is coupled to $\myhat{\mv{J}}$ yielding the total angular momentum $\myhat{\mv{F}}$: $\myhat{\mv{F}} = \myhat{\mv{J}} + \myhat{\mv{I}}$. The basis functions are denoted by $ \Ket{n \Lambda, v, N \Lambda S J I F \MF}$ and $\Ket{\np \Lp, \vp, \Np \Lp \Sp \Jp \Ip \Fp \MFp}$ for the neutral and ionic states, respectively. Here, $I$ and $F$ ($\Ip$ and $\Fp$)  denote the nuclear spin and the total angular momentum quantum number, respectively, of the neutral (ionic) state. $\MF$ and $\MFp$ denote the projection angular momentum quantum numbers with respect to the space-fixed $z$-axis associated with $F$ and $\Fp$. All other quantum numbers are defined as before, see Tab.~\ref{tab:QNs_direct_ionization}. The photoionization transition probability is then proportional to the quantity 
\begin{widetext}
\begin{align}
P(F,\Fp) = \sum_{l=0}^{\infty} \sum_{m_l=-l}^{l}  \sum_{m_s=-s}^{s} \sum_{\MF=-F}^{F}\sum_{\MFp=-\Fp}^{\Fp} 
\Big | & 
\left (
\Bra{\np \Lp, \vp, \Np \Lp \Sp \Jp \Ip \Fp \MFp}  \Bra{s,m_s}  \Bra{l,m_l}
 \right )
\nonumber \\ &
\boldsymbol{\myhat\mu} 
 \Ket{n \Lambda, v, N \Lambda S J I F \MF} \Big  |^2.
 \label{eq_hfs_transition_probability}
\end{align}
\end{widetext}

The coupled angular momentum states are expressed in the decoupled tensor-product basis of the spin-rotational-vibronic and the nuclear spin states as
\begin{multline}
\ket{n \Lambda, v, N \Lambda S J I F \MF} = \\
 \sum_{\MI}\sum_{\MJ} \CG{J}{\MJ}{I}{\MI}{F}{\MF} \ket{n \Lambda, v, N \Lambda S J \MJ, I \MI},
\end{multline}
and equivalently for the ionic state. In this basis, the transition matrix element appearing in Eq.~\eqref{eq_hfs_transition_probability} accounts for
\begin{align}
& 
\left (
\Bra{\np \Lp, \vp, \Np \Lp \Sp \Jp \Ip \Fp \MFp} \Bra{s,m_s}  \Bra{l,m_l} 
\right )
 \nonumber \\
& \quad \times \boldsymbol{\myhat\mu}  \Ket{n \Lambda, v, N \Lambda S J I F \MF} = \nonumber \\ 
&\quad\sum_{\MIp,\MI}\sum_{\MJp,\MJ} \CG{\Jp}{\MJp}{\Ip}{\MIp}{\Fp}{\MFp} \CG{J}{\MJ}{I}{\MI}{F}{\MF} \nonumber \\ 
&\quad 
\times
 (
 \bra{\np \Lp, \vp, \Np \Lp \Sp \Jp \MJp, \Ip \MIp} \Bra{s,m_s}  
\nonumber \\ 
&\quad \times \Bra{l,m_l} 
) \boldsymbol{\myhat\mu} \ket{n \Lambda, v, N \Lambda S J \MJ, I \MI}.
\end{align}
Since the nuclear spin is neither affected by the absorption of electromagnetic radiation, nor by the ejection of the photoelectron \cite{signorell97c}, we may separate the nuclear spin states from the remaining transition matrix element obtaining
\begin{align}
&
\left (
\Bra{\np \Lp, \vp, \Np \Lp \Sp \Jp \MJp, \Ip \MIp} \Bra{s,m_s}  \Bra{l,m_l} 
\right )
\nonumber \\
&  \times \boldsymbol{\myhat\mu}  \Ket{n \Lambda, v, N \Lambda S J \MJ, I \MI} =
\nonumber \\ 
&  \kd{\Ip}{I} \kd{\MIp}{\MI}
\big (
\Bra{\np \Lp, \vp, \Np \Lp \Sp \Jp \MJp} 
\nonumber \\
& \times
\Bra{s,m_s}  \Bra{l,m_l} \big ) 
\boldsymbol{\myhat\mu} 
 \Ket{n \Lambda, v, N \Lambda S J \MJ}.
\label{eq_hfs_trans_mat_el_I_separated}
\end{align}
The transition matrix element on the last and next-to-last line of Eq.~\eqref{eq_hfs_trans_mat_el_I_separated} is the same as in Eq.~\eqref{eq_fs_complete_trans_mat_el_with_9j}. Substituting Eq.~\eqref{eq_fs_complete_trans_mat_el_with_9j} into \eqref{eq_hfs_trans_mat_el_I_separated}, replacing Clebsch-Gordan coefficients by 3j-symbols and simplifying the result \cite{germann16a} yields the following expression for the transition probability:
\begin{align}
&P(F,\Fp) = \nonumber \\
&\quad  (2 \Np +1) (2N+1)  (2 S +1)  (2 \Jp +1) (2J+1)  \nonumber \\ 
&\quad \times (2 \Fp +1) (2F+1) \kd{I}{\Ip}  \sum _l  \sum _{k=l\pm 1} (2k+1)  \nonumber \\ 
&\quad \times\Big|\Bra{\np \Lp, \vp} \formsphtMolFixed{k}{\Delta\Lambda} \Ket{n \Lambda, v}\Big|^{2} \tj{\Np}{k}{N}{-\Lp}{\Delta\Lambda}{\Lambda}^{2} \nonumber \\
&\quad\times \sum_{u=|k-s|}^{k+s} (2u+1) \ninej{\Jp}{u}{J}{\Np}{k}{N}{\Sp}{s}{S}^{2} \sj{u}{J}{\Jp}{I}{\Fp}{F}^{2}\nonumber \\
&\quad\times \sum _{m_l}\tj{l}{ 1}{k}{ -m_l}{ \mu_0}{ -p}^2  \sum _{m_s} \tj{u}{k}{s}{\upr}{-p}{m_s}^2.
\label{eq_hfs_transition_probability_simplified}
\end{align}
The terms on the last line above again account for $1/(3(2k+1))$ for linear polarized radiation ($ \mu_0 = 0 $), i.e.,
\begin{align}
& P(F,\Fp) = \nonumber \\
&\quad \frac{1}{3} (2 \Np +1) (2N+1)  (2 S +1)  (2 \Jp +1) (2J+1) \nonumber \\
& \quad \times (2 \Fp +1) (2F+1) \kd{I}{\Ip} \sum _l  \sum _{k=l\pm 1} \nonumber \\
& \quad \times \left |\Braket{\np \Lp, \vp | \formsphtMolFixed{k}{\Delta\Lambda} | n \Lambda, v}\right |^{2} \tj{\Np}{k}{N}{-\Lp}{\Delta\Lambda}{\Lambda}^{2} \nonumber \\ 
&\quad \times \sum_{u=|k-s|}^{k+s} (2u+1)  \ninej{\Jp}{u}{J}{\Np}{k}{N}{\Sp}{s}{S}^{2} \sj{u}{J}{\Jp}{I}{\Fp}{F}^{2}.
\label{eq_hfs_transition_probability_lin_pol}
\end{align}

In essence, we have reproduced in Eq.~\eqref{eq_hfs_transition_probability_lin_pol} the result from Eq.~\eqref{eq_fs_transition_probability_lin_pol} with an additional Wigner 6j-symbol describing the influence of the nuclear spin, i.e., the hyperfine structure effects. Since the photoelectron does not carry nuclear spin, the same selection rule as for $\Delta J$ applies also for $\Delta F$, namely $| \Delta F | = | \Fp - F | \le u$. Again due to the separability of the transition matrix element, the relative intensities of photoionizing transitions between particular hyperfine levels are determined by the magnitude of the vibronic transition matrix elements which also determine the intensities of different rotational lines in a photoelectron spectrum. For vanishing nuclear spin, Eq.~\eqref{eq_hfs_transition_probability_lin_pol} reduces to Eq.~\eqref{eq_fs_transition_probability_lin_pol}.

\section{Applications}

\subsection{Example 1: Fine-structure resolved photoionization of molecular oxygen}

In order to validate the present results, Eq.~\eqref{eq_fs_transition_probability_lin_pol} is applied to the analysis of the spin-rotation-resolved photoelectron spectrum of the \OtwoXstate $\rightarrow$ \OtwoPlusbstate photoionizing transition in molecular oxygen reported by Palm and Merkt \cite{palm98c}.

\begin{figure}[htbp]
\begin{center}
\includegraphics{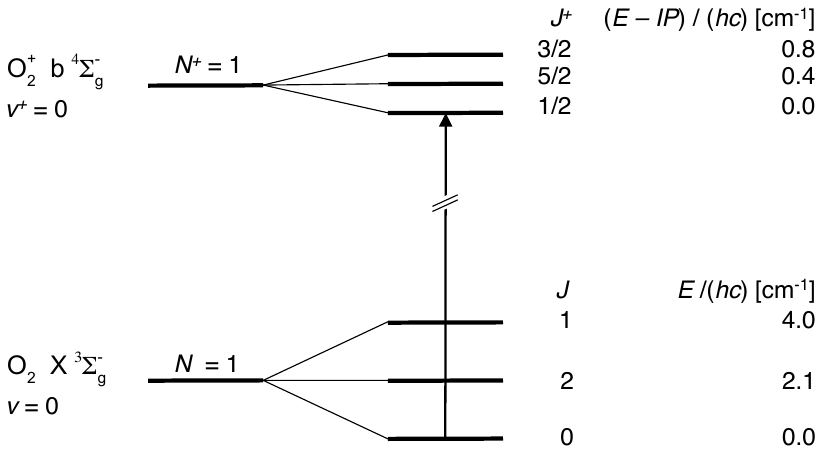}
\caption{Spin-rotational energy-level structure of the \OtwoPlusbstate and the \OtwoXstate states of the O$_2^+$ ion and of neutral O$_2$, respectively. Levels are labeled by their total angular momentum quantum numbers $J$ and $\Jp$ for the neutral and ionic state, respectively, as well as their relative term value (adapted from Ref. \onlinecite{palm98c}).}
\label{fig:O2_level_scheme}
\end{center}
\end{figure}

The energy level structure of neutral O$_2$ in the electronic ground state \OtwoXstate and of the \OtwoPlus ion in the \OtwoPlusbstate state is shown in Fig.~\ref{fig:O2_level_scheme}. We are interested in the Q(1) line of the $v=0 \rightarrow \vp =0$ band, i.e., in the transition $v=0, N=1 \rightarrow \vp=0, \Np=1$, which has been measured with the highest resolution.
Neutral O$_2$ exhibits a total electron spin $S = 1$ in the \OtwoXstate state, such that there are three spin-rotation components for $N=1$: $J=0$, 1 and 2. For O$_2^+$ in the \OtwoPlusbstate state, the total electronic spin is $\Sp = 3/2$ giving rise to three spin-rotation components with $\Jp = 1/2$, 3/2 and 5/2.
Hence, there are in total nine different transitions between the fine-structure components of the neutral and ionic state involved in the Q(1) line.

\begin{figure}[!h]
\begin{center}
\includegraphics[scale=\mathematicaFig]{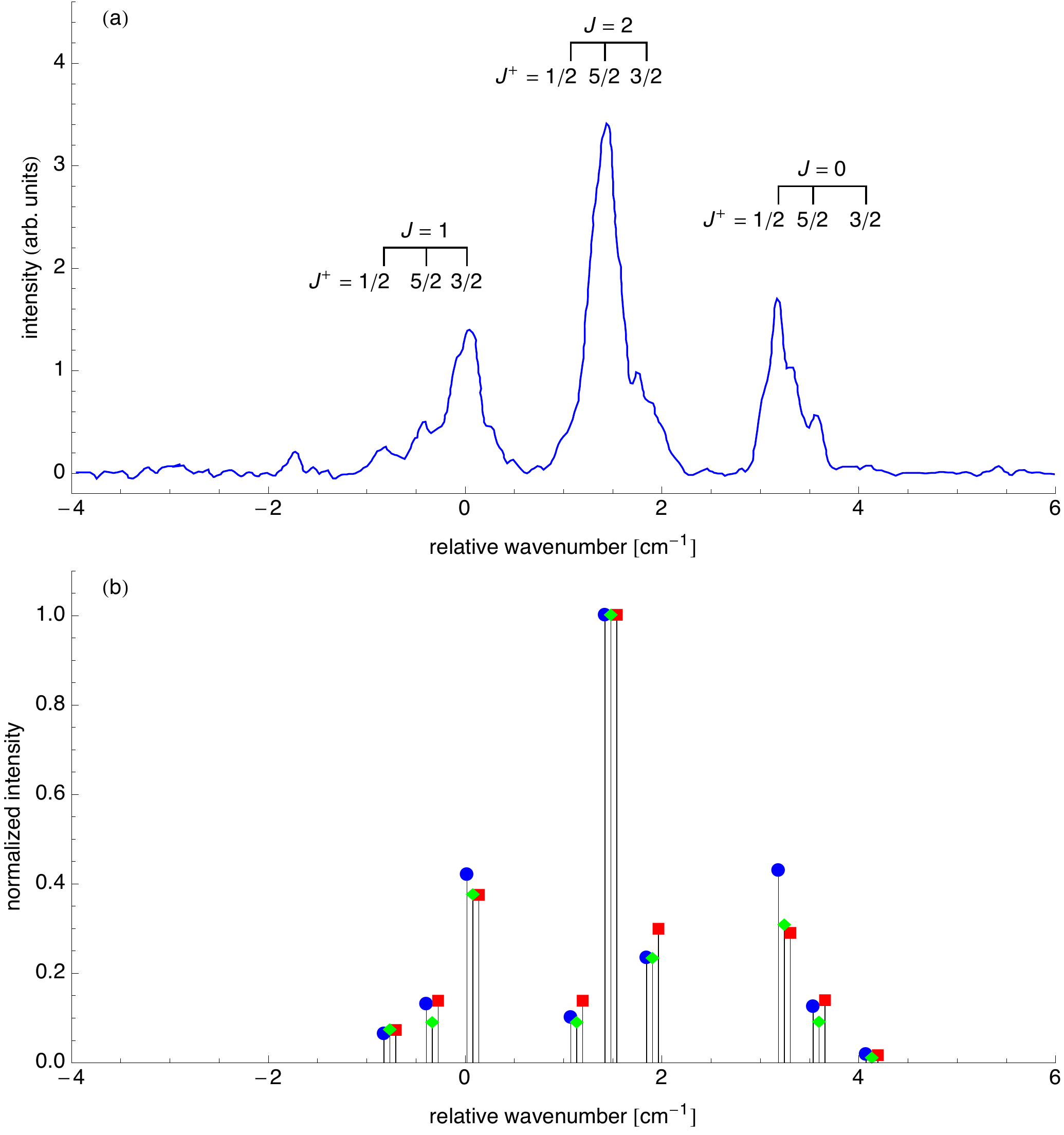}
\caption{
(a) Measured fine-structure-resolved photoionization spectrum of the O$_2$~\OtwoXstate $(v=0, N=1) \rightarrow$ \OtwoPlus~\OtwoPlusbstate $(\vp=0, \Np=1)$ transition recorded by Palm and Merkt, digitized from Fig.~4 of Ref.~\onlinecite{palm98c}. The assignment bars indicate transitions between levels of total angular momentum quantum number $J$ and $J^+$ in the neutral and ionic states, respectively.
(b) Stick spectrum showing the normalized intensities of individual spin-rotation-resolved transitions. Blue circles: experimental intensities extracted from the spectrum in (a). Green diamonds: fit of our photoionization model Eq.~\eqref{eq_fs_transition_probability_lin_pol_singlet} to the measured intensities with two vibronic coefficients $C_0$ and $C_2$ treated as free parameters. Red squares: spin-rotation-resolved intensities calculated with Eq.~\eqref{eq_fs_transition_probability_lin_pol_singlet} using the relative values for the vibronic coefficients $C_0=0.6$ and $C_2=0.4$ determined from a rotationally-resolved photoelectron spectrum in Ref. \onlinecite{hsu98a}. All intensities are normalized to unity for the most intense transition.)
}
\label{fig:O2_spectrum}
\end{center}
\end{figure}

The experimental photoelectron spectrum of the fine-structure-resolved Q(1) line of Palm and Merkt \cite{palm98c} is reproduced in Fig.~\ref{fig:O2_spectrum}~(a). The spectrum shows three well-separated peaks spaced by about $\SI{2}{\per\centi\meter}$ reflecting the spin-rotation splitting in the \OtwoXstate state of neutral O$_2$. Every peak is composed of three partially overlapping lines which stem from transitions to different spin-rotation components of the ionic \OtwoPlusbstate state which are spaced by about $\SI{0.4}{\per\centi\meter}$, totaling to the nine spin-rotation transitions of the Q(1) line.

To compare the measured spectrum with our theoretical predictions, the experimental line intensities have been extracted from the measured spectrum by fitting Gaussian functions to the spectral features. The empirical intensities found this way are shown as blue circles in the stick spectrum of Fig.~\ref{fig:O2_spectrum}~(b).

From the analysis of  rotationally, but not fine-structure-resolved photoelectron spectra of Hsu et al.\cite{hsu98a}, it has been established that the photoionization of O$_2$ mainly involves the coefficient $C_0$ and to a smaller extent  $C_2$. For $k>2$, the vibronic matrix elements essentially vanish.

According to our model, for $k=0$ we have $u=1/2$ in Eq.~\eqref{eq_fs_transition_probability_lin_pol}, giving rise to transitions with $\Delta J = \pm 1/2$ with high intensities. Indeed, the lines with the highest intensity within each of three peaks for $J=0$, 1 and 2 in Fig.~\ref{fig:O2_spectrum} (a) obey this criterion. On the contrary, transitions with $| \Delta J| > 1/2$ are only possible via the $k = 2$ vibronic matrix element with a considerably reduced magnitude. Indeed, such lines show only low to medium intensities in the experimental spectrum.

For a quantitative analysis, we fitted the normalized experimental intensities shown as blue circles in the stick spectrum of Fig.~\ref{fig:O2_spectrum}~(b) to the relative ionization rates as given by the transition probabilities from Eq.~\eqref{eq_fs_transition_probability_lin_pol} weighted by the neutral-state level populations. The latter were calculated from a Boltzmann distribution with a rotational temperature of $\SI{7}{\kelvin}$ as reported in Ref. \onlinecite{palm98c}. The two vibronic coefficients $C_0$ and $C_2$ were treated as free parameters. The line intensities obtained from this procedure are shown as green diamonds in Fig.~\ref{fig:O2_spectrum}~(b). As can be seen, the present model reproduces the measured photoionization intensities well. The ratio of the two vibronic coefficients which accounts for the relative intensities of the $k=0$ vs. $k=2$ line obtained from the fit amounts to $C_2/C_0 = 0.3/0.7$. This result is in agreement with the values of $C_2 = 0.4 \pm 0.1$, $C_0 = 0.6 \pm 0.1$ found in Ref.~\onlinecite{hsu98a} in the analysis of a rotationally---instead of fine-structure---resolved photoelectron spectrum. For comparison, the red squares in Fig.~\ref{fig:O2_spectrum} (b) show the predicted intensities using the values of the vibronic coefficients from Ref. \onlinecite{hsu98a}. In this approach, no free parameters except a global intensity normalization factor enter our modeling. Also in this case, the agreement with experiment is very good.


\subsection{Example 2: Hyperfine propensities in the photoionization of molecular nitrogen}

\begin{figure}[!b]
\begin{center}
\includegraphics{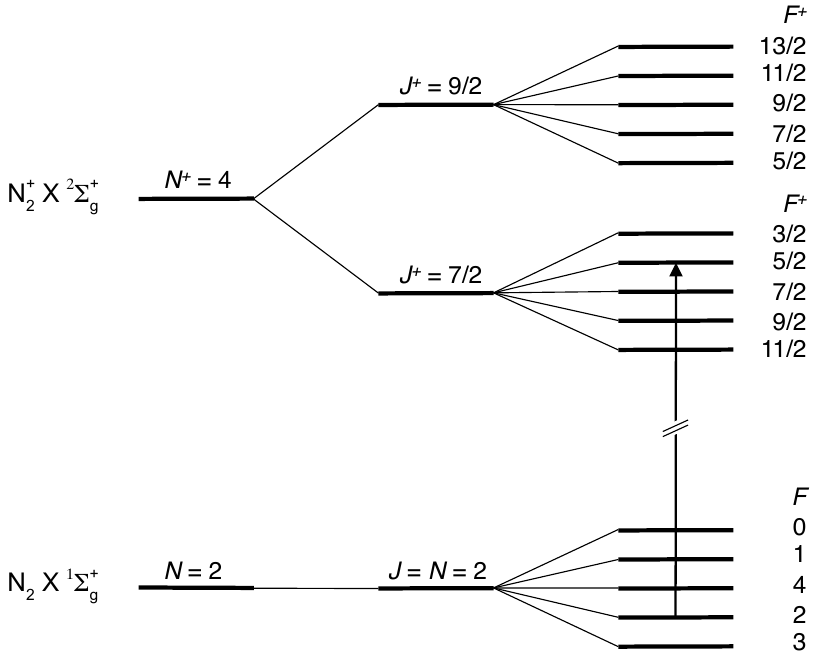}
\caption{Energy-level structure of the nitrogen molecular ion \cite{rosner85a} and the neutral nitrogen molecule. The energetic order of the hyperfine levels in the neutral N$_2$ $^1\Sigma_\mathrm{g}^+$ state has been estimated using electric-quadrupole coupling constants extrapolated from spectroscopic data on the neutral N$_2$ A$^3\Sigma^+_\mathrm u$ state and from N$_2$ complexes \cite{freund70a,desantis73a,legon92a}. }
\label{fig:N2+_one_photon_ionization_level_scheme}
\end{center}
\end{figure}

We are not aware of any hyperfine-resolved photoelectron spectra reported in the literature that could serve as a benchmark to test our photoionization model. In order to illustrate the implications of hyperfine structure in direct photoionization, we calculated the relative intensities of hyperfine-resolved photoionization transitions of molecular nitrogen which govern the hyperfine populations of the resulting cations \cite{germann14a}. As an illustrative example, we studied the intensities of hyperfine components of the transition N$_2$ X$^1\Sigma_\mathrm{g}^+$ $N=2$ $\rightarrow$ \ntp \ntwopXstate $\Np=4$. A level scheme is shown in Fig.~\ref{fig:N2+_one_photon_ionization_level_scheme}. As the neutral electronic ground state of N$_2$ is a singlet state, the total electron spin vanishes and we have $N=J$. The $^{14}$N isotope has a nuclear spin of 1. For N$_2$, a total nuclear spin of $I=0$ or 2 is possible for $N=2$ according to the Pauli principle. We concentrate on the case $I=2$. The \ntp ion exhibits fine and hyperfine structure. The rotational levels are split by the spin-rotation interaction into two spin-rotation components labeled by the quantum number $J$, which may take the values $\Jp=\Np+1/2$ and $\Jp=\Np-1/2$. The spin-rotation levels are split further into hyperfine levels associated with the quantum number $\Fp$ with the values $\Fp=\Jp+\Ip, \Jp+\Ip - 1,\dots , |\Jp-\Ip|$.

Similar to the previous example of O$_2$, the N$_2$ X$^1\Sigma_\mathrm{g}^+$ $\rightarrow$ \ntp \ntwopXstate photoionization transition  is dominated by the vibronic coefficients $C_0$ and $C_2$ \cite{baltzer92a,oehrwall99a}. Coefficients with $k>2$ are considerably smaller and are thus neglected here. For the $N=2 \rightarrow \Np=4$ transitions, we have $J=N=2$ and $\Jp= 9/2$ or 7/2. Since for both of these transitions we have $|\Delta J| > 1/2$, they may solely occur via the  vibronic transition matrix element associated with $C_2$.

\begin{figure}[tpb]
\begin{center}
\includegraphics[scale=\mathematicaFig]{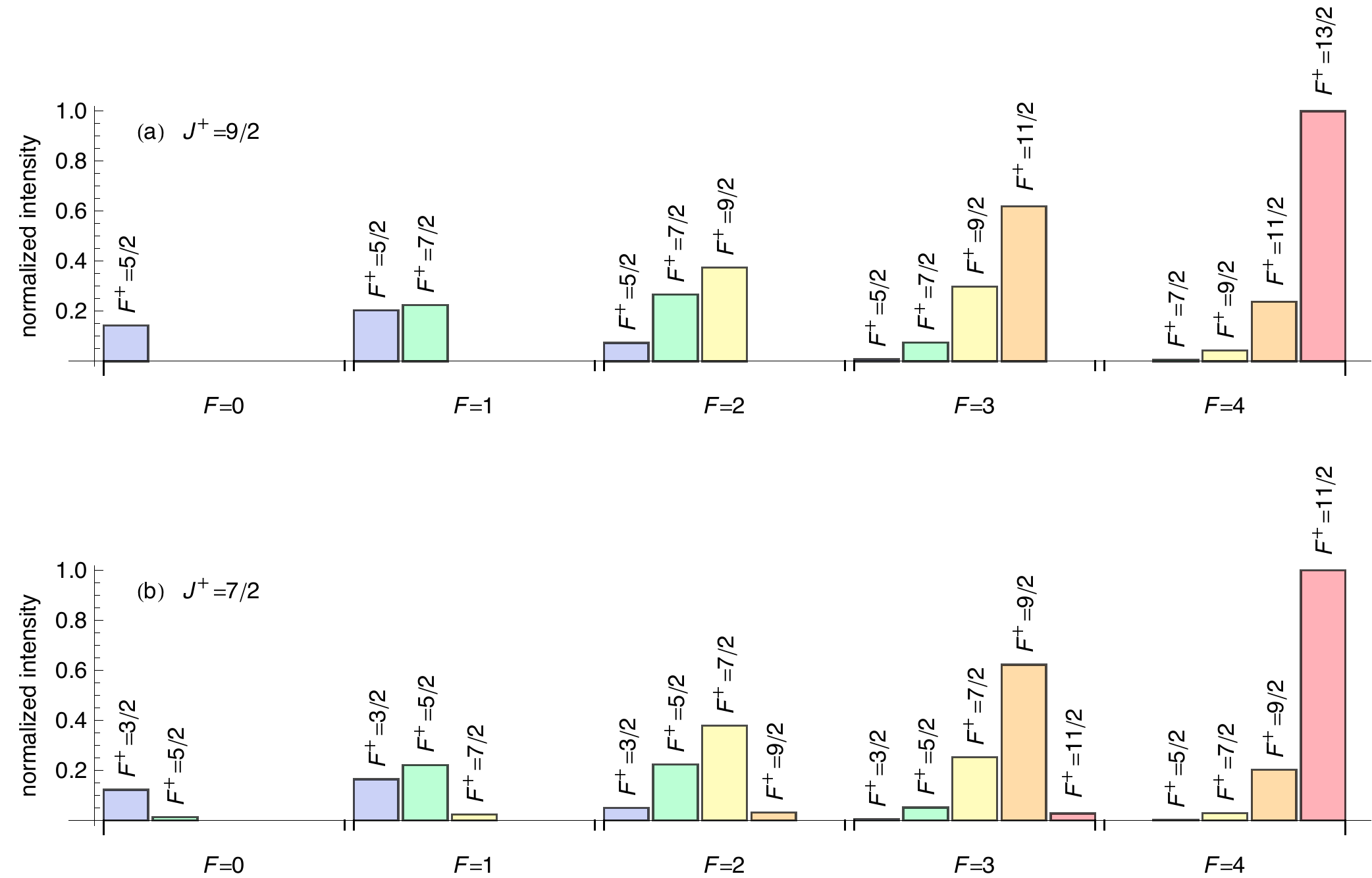}
\caption{
Relative intensities of hyperfine components of the N$_2$ X$^1\Sigma_\mathrm{g}^+$ $N=J=2 \rightarrow$ \ntp \ntwopXstate $\Np=4, \Jp=9/2$ (a) and $\Jp=7/2$ (b)  photoionization transition calculated by Eq.~\eqref{eq_hfs_transition_probability_lin_pol}. The hyperfine levels in the neutral and ionic state are labeled by their total angular momentum quantum number $F$ and $\Fp$, respectively. Transitions with relative intensities $< 10^{-3}$ have been suppressed. }
\label{fig:N2_ionisation_intensities}
\end{center}
\end{figure}

The intensities of the hyperfine components calculated from Eq.~\eqref{eq_hfs_transition_probability_lin_pol} are shown in Fig.~\ref{fig:N2_ionisation_intensities}. The relative intensities of the hyperfine transitions are governed by the 6j-symbol on the last line of Eq.~\eqref{eq_hfs_transition_probability_lin_pol}. A propensity towards transitions obeying the relation $\Delta J = \Delta F$ is observed, similar as has previously been observed in bound-bound hyperfine-resolved transitions \cite{rosner85a, germann16b}.

\section{Summary and conclusions}

In the present paper, we have developed a general framework for fine- and hyperfine intensities in the direct photoionization of molecules. We have derived closed, symmetrized expressions of the relevant squared photoionization-transition matrix elements which lend themselves to a straightforward derivation of photoionization selection rules and to an efficient computational implementation. The present results can be used in the analysis of fine- and hyperfine resolved photoelectron spectra and the prediction of cationic level populations upon photoionization. 

The present treatment does not, however, contain the effects of interactions between different ionization channels \cite{merkt93c}. Whereas these effects often play only a minor role in direct photoionization, they may be observed in Rydberg and pulsed-field-ionization zero-kinetic-energy (PFI-ZEKE) photoelectron spectra. In this case, a more detailed treatment of the scattering problem of the photoelectron within the framework of MQDT is warranted \cite{woerner03a,osterwalder04a,cruse08a,jungen11a,sprecher14a}.


\begin{acknowledgments}
This work has been supported by the Swiss National Science Foundation as part of the National Centre of Competence in Research, Quantum Science \& Technology (NCCR-QSIT), the European Commission under the Seventh Framework Programme FP7 GA 607491 COMIQ and the University of Basel
\end{acknowledgments}

\bibliography{Main-Jan2016_modified,HF_PI_dynamics_refs_reduced_arXiv}

\begin{thebibliography}{40}%
\makeatletter
\providecommand \@ifxundefined [1]{%
 \@ifx{#1\undefined}
}%
\providecommand \@ifnum [1]{%
 \ifnum #1\expandafter \@firstoftwo
 \else \expandafter \@secondoftwo
 \fi
}%
\providecommand \@ifx [1]{%
 \ifx #1\expandafter \@firstoftwo
 \else \expandafter \@secondoftwo
 \fi
}%
\providecommand \natexlab [1]{#1}%
\providecommand \enquote  [1]{``#1''}%
\providecommand \bibnamefont  [1]{#1}%
\providecommand \bibfnamefont [1]{#1}%
\providecommand \citenamefont [1]{#1}%
\providecommand \href@noop [0]{\@secondoftwo}%
\providecommand \href [0]{\begingroup \@sanitize@url \@href}%
\providecommand \@href[1]{\@@startlink{#1}\@@href}%
\providecommand \@@href[1]{\endgroup#1\@@endlink}%
\providecommand \@sanitize@url [0]{\catcode `\\12\catcode `\$12\catcode
  `\&12\catcode `\#12\catcode `\^12\catcode `\_12\catcode `\%12\relax}%
\providecommand \@@startlink[1]{}%
\providecommand \@@endlink[0]{}%
\providecommand \url  [0]{\begingroup\@sanitize@url \@url }%
\providecommand \@url [1]{\endgroup\@href {#1}{\urlprefix }}%
\providecommand \urlprefix  [0]{URL }%
\providecommand \Eprint [0]{\href }%
\providecommand \doibase [0]{http://dx.doi.org/}%
\providecommand \selectlanguage [0]{\@gobble}%
\providecommand \bibinfo  [0]{\@secondoftwo}%
\providecommand \bibfield  [0]{\@secondoftwo}%
\providecommand \translation [1]{[#1]}%
\providecommand \BibitemOpen [0]{}%
\providecommand \bibitemStop [0]{}%
\providecommand \bibitemNoStop [0]{.\EOS\space}%
\providecommand \EOS [0]{\spacefactor3000\relax}%
\providecommand \BibitemShut  [1]{\csname bibitem#1\endcsname}%
\let\auto@bib@innerbib\@empty
\bibitem [{\citenamefont {Turner}\ \emph {et~al.}(1970)\citenamefont {Turner},
  \citenamefont {Baker}, \citenamefont {Baker},\ and\ \citenamefont
  {Brundle}}]{turner70a}%
  \BibitemOpen
  \bibfield  {author} {\bibinfo {author} {\bibfnamefont {D.~W.}\ \bibnamefont
  {Turner}}, \bibinfo {author} {\bibfnamefont {C.}~\bibnamefont {Baker}},
  \bibinfo {author} {\bibfnamefont {A.~D.}\ \bibnamefont {Baker}}, \ and\
  \bibinfo {author} {\bibfnamefont {C.~R.}\ \bibnamefont {Brundle}},\
  }\href@noop {} {\emph {\bibinfo {title} {Molecular Photoelectron
  Spectroscopy.}}}\ (\bibinfo  {publisher} {Wiley-Interscience},\ \bibinfo
  {address} {London},\ \bibinfo {year} {1970})\BibitemShut {NoStop}%
\bibitem [{\citenamefont {Ellis}, \citenamefont {Feher},\ and\ \citenamefont
  {Wright}(2005)}]{ellis05a}%
  \BibitemOpen
  \bibfield  {author} {\bibinfo {author} {\bibfnamefont {A.~M.}\ \bibnamefont
  {Ellis}}, \bibinfo {author} {\bibfnamefont {M.}~\bibnamefont {Feher}}, \ and\
  \bibinfo {author} {\bibfnamefont {T.~G.}\ \bibnamefont {Wright}},\
  }\href@noop {} {\emph {\bibinfo {title} {Electronic and Photoelectron
  Spectroscopy}}}\ (\bibinfo  {publisher} {Cambridge University Press},\
  \bibinfo {address} {Cambridge},\ \bibinfo {year} {2005})\BibitemShut
  {NoStop}%
\bibitem [{\citenamefont {Merkt}, \citenamefont {Willitsch},\ and\
  \citenamefont {Hollenstein}(2011)}]{merkt11a}%
  \BibitemOpen
  \bibfield  {author} {\bibinfo {author} {\bibfnamefont {F.}~\bibnamefont
  {Merkt}}, \bibinfo {author} {\bibfnamefont {S.}~\bibnamefont {Willitsch}}, \
  and\ \bibinfo {author} {\bibfnamefont {U.}~\bibnamefont {Hollenstein}},\
  }\bibfield  {title} {\enquote {\bibinfo {title} {{High-resolution
  Photoelectron Spectroscopy}},}\ }in\ \href@noop {} {\emph {\bibinfo
  {booktitle} {Handbook of High-resolution Spectroscopy}}},\ Vol.~\bibinfo
  {volume} {3},\ \bibinfo {editor} {edited by\ \bibinfo {editor} {\bibfnamefont
  {M.}~\bibnamefont {Quack}}\ and\ \bibinfo {editor} {\bibfnamefont
  {F.}~\bibnamefont {Merkt}}}\ (\bibinfo  {publisher} {John Wiley \& Sons},\
  \bibinfo {address} {Hoboken},\ \bibinfo {year} {2011})\ p.\ \bibinfo {pages}
  {1617}\BibitemShut {NoStop}%
\bibitem [{\citenamefont {Willitsch}\ and\ \citenamefont
  {Merkt}(2005)}]{willitsch05b}%
  \BibitemOpen
  \bibfield  {author} {\bibinfo {author} {\bibfnamefont {S.}~\bibnamefont
  {Willitsch}}\ and\ \bibinfo {author} {\bibfnamefont {F.}~\bibnamefont
  {Merkt}},\ }\href@noop {} {\bibfield  {journal} {\bibinfo  {journal} {Int. J.
  Mass Spectrom.}\ }\textbf {\bibinfo {volume} {245}},\ \bibinfo {pages} {14}
  (\bibinfo {year} {2005})}\BibitemShut {NoStop}%
\bibitem [{\citenamefont {Buckingham}, \citenamefont {Orr},\ and\ \citenamefont
  {Sichel}(1970)}]{buckingham70a}%
  \BibitemOpen
  \bibfield  {author} {\bibinfo {author} {\bibfnamefont {A.~D.}\ \bibnamefont
  {Buckingham}}, \bibinfo {author} {\bibfnamefont {B.~J.}\ \bibnamefont {Orr}},
  \ and\ \bibinfo {author} {\bibfnamefont {J.~M.}\ \bibnamefont {Sichel}},\
  }\href@noop {} {\bibfield  {journal} {\bibinfo  {journal} {Phil. Trans. Roy.
  Soc. Lond. A}\ }\textbf {\bibinfo {volume} {268}},\ \bibinfo {pages} {147}
  (\bibinfo {year} {1970})}\BibitemShut {NoStop}%
\bibitem [{\citenamefont {Dixit}\ and\ \citenamefont {McKoy}(1985)}]{dixit85a}%
  \BibitemOpen
  \bibfield  {author} {\bibinfo {author} {\bibfnamefont {S.~N.}\ \bibnamefont
  {Dixit}}\ and\ \bibinfo {author} {\bibfnamefont {V.}~\bibnamefont {McKoy}},\
  }\href@noop {} {\bibfield  {journal} {\bibinfo  {journal} {J.~Chem. Phys.}\
  }\textbf {\bibinfo {volume} {82}},\ \bibinfo {pages} {3546} (\bibinfo {year}
  {1985})}\BibitemShut {NoStop}%
\bibitem [{\citenamefont {Allendorf}\ \emph {et~al.}(1989)\citenamefont
  {Allendorf}, \citenamefont {Leahy}, \citenamefont {Jacobs},\ and\
  \citenamefont {Zare}}]{allendorf89a}%
  \BibitemOpen
  \bibfield  {author} {\bibinfo {author} {\bibfnamefont {S.~W.}\ \bibnamefont
  {Allendorf}}, \bibinfo {author} {\bibfnamefont {D.~J.}\ \bibnamefont
  {Leahy}}, \bibinfo {author} {\bibfnamefont {D.~C.}\ \bibnamefont {Jacobs}}, \
  and\ \bibinfo {author} {\bibfnamefont {R.~N.}\ \bibnamefont {Zare}},\
  }\href@noop {} {\bibfield  {journal} {\bibinfo  {journal} {J. Chem. Phys.}\
  }\textbf {\bibinfo {volume} {91}},\ \bibinfo {pages} {2216} (\bibinfo {year}
  {1989})}\BibitemShut {NoStop}%
\bibitem [{\citenamefont {Wang}\ and\ \citenamefont {McKoy}(1991)}]{wang91a}%
  \BibitemOpen
  \bibfield  {author} {\bibinfo {author} {\bibfnamefont {K.}~\bibnamefont
  {Wang}}\ and\ \bibinfo {author} {\bibfnamefont {V.}~\bibnamefont {McKoy}},\
  }\href@noop {} {\bibfield  {journal} {\bibinfo  {journal} {J.~Chem. Phys.}\
  }\textbf {\bibinfo {volume} {95}},\ \bibinfo {pages} {4977} (\bibinfo {year}
  {1991})}\BibitemShut {NoStop}%
\bibitem [{\citenamefont {Xie}\ and\ \citenamefont {Zare}(1992)}]{xie92a}%
  \BibitemOpen
  \bibfield  {author} {\bibinfo {author} {\bibfnamefont {J.}~\bibnamefont
  {Xie}}\ and\ \bibinfo {author} {\bibfnamefont {R.~N.}\ \bibnamefont {Zare}},\
  }\href@noop {} {\bibfield  {journal} {\bibinfo  {journal} {J.~Chem. Phys.}\
  }\textbf {\bibinfo {volume} {97}},\ \bibinfo {pages} {2891} (\bibinfo {year}
  {1992})}\BibitemShut {NoStop}%
\bibitem [{\citenamefont {Dixit}\ and\ \citenamefont {McKoy}(1986)}]{dixit86a}%
  \BibitemOpen
  \bibfield  {author} {\bibinfo {author} {\bibfnamefont {S.~N.}\ \bibnamefont
  {Dixit}}\ and\ \bibinfo {author} {\bibfnamefont {V.}~\bibnamefont {McKoy}},\
  }\href@noop {} {\bibfield  {journal} {\bibinfo  {journal} {Chem. Phys.
  Lett.}\ }\textbf {\bibinfo {volume} {128}},\ \bibinfo {pages} {49} (\bibinfo
  {year} {1986})}\BibitemShut {NoStop}%
\bibitem [{\citenamefont {Xie}\ and\ \citenamefont {Zare}(1990)}]{xie90a}%
  \BibitemOpen
  \bibfield  {author} {\bibinfo {author} {\bibfnamefont {J.}~\bibnamefont
  {Xie}}\ and\ \bibinfo {author} {\bibfnamefont {R.~N.}\ \bibnamefont {Zare}},\
  }\href@noop {} {\bibfield  {journal} {\bibinfo  {journal} {J.~Chem. Phys.}\
  }\textbf {\bibinfo {volume} {93}},\ \bibinfo {pages} {3033} (\bibinfo {year}
  {1990})}\BibitemShut {NoStop}%
\bibitem [{\citenamefont {Signorell}\ and\ \citenamefont
  {Merkt}(1997)}]{signorell97c}%
  \BibitemOpen
  \bibfield  {author} {\bibinfo {author} {\bibfnamefont {R.}~\bibnamefont
  {Signorell}}\ and\ \bibinfo {author} {\bibfnamefont {F.}~\bibnamefont
  {Merkt}},\ }\href@noop {} {\bibfield  {journal} {\bibinfo  {journal} {Mol.
  Phys.}\ }\textbf {\bibinfo {volume} {92}},\ \bibinfo {pages} {793} (\bibinfo
  {year} {1997})}\BibitemShut {NoStop}%
\bibitem [{\citenamefont {Palm}\ and\ \citenamefont {Merkt}(1998)}]{palm98c}%
  \BibitemOpen
  \bibfield  {author} {\bibinfo {author} {\bibfnamefont {H.}~\bibnamefont
  {Palm}}\ and\ \bibinfo {author} {\bibfnamefont {F.}~\bibnamefont {Merkt}},\
  }\href@noop {} {\bibfield  {journal} {\bibinfo  {journal} {Phys. Rev. Lett.}\
  }\textbf {\bibinfo {volume} {81}},\ \bibinfo {pages} {1385} (\bibinfo {year}
  {1998})}\BibitemShut {NoStop}%
\bibitem [{\citenamefont {W\"{o}rner}, \citenamefont {Hollenstein},\ and\
  \citenamefont {Merkt}(2003)}]{woerner03a}%
  \BibitemOpen
  \bibfield  {author} {\bibinfo {author} {\bibfnamefont {H.~J.}\ \bibnamefont
  {W\"{o}rner}}, \bibinfo {author} {\bibfnamefont {U.}~\bibnamefont
  {Hollenstein}}, \ and\ \bibinfo {author} {\bibfnamefont {F.}~\bibnamefont
  {Merkt}},\ }\href@noop {} {\bibfield  {journal} {\bibinfo  {journal} {Phys.
  Rev.~A}\ }\textbf {\bibinfo {volume} {68}},\ \bibinfo {pages} {032510}
  (\bibinfo {year} {2003})}\BibitemShut {NoStop}%
\bibitem [{\citenamefont {Osterwalder}\ \emph {et~al.}(2004)\citenamefont
  {Osterwalder}, \citenamefont {W\"{u}est}, \citenamefont {Merkt},\ and\
  \citenamefont {Jungen}}]{osterwalder04a}%
  \BibitemOpen
  \bibfield  {author} {\bibinfo {author} {\bibfnamefont {A.}~\bibnamefont
  {Osterwalder}}, \bibinfo {author} {\bibfnamefont {A.}~\bibnamefont
  {W\"{u}est}}, \bibinfo {author} {\bibfnamefont {F.}~\bibnamefont {Merkt}}, \
  and\ \bibinfo {author} {\bibfnamefont {{\relax Ch}.}~\bibnamefont {Jungen}},\
  }\href@noop {} {\bibfield  {journal} {\bibinfo  {journal} {J.~Chem. Phys.}\
  }\textbf {\bibinfo {volume} {121}},\ \bibinfo {pages} {11810} (\bibinfo
  {year} {2004})}\BibitemShut {NoStop}%
\bibitem [{\citenamefont {Cruse}, \citenamefont {\mbox{Ch.} Jungen},\ and\
  \citenamefont {Merkt}(2008)}]{cruse08a}%
  \BibitemOpen
  \bibfield  {author} {\bibinfo {author} {\bibfnamefont {H.~A.}\ \bibnamefont
  {Cruse}}, \bibinfo {author} {\bibnamefont {\mbox{Ch.} Jungen}}, \ and\
  \bibinfo {author} {\bibfnamefont {F.}~\bibnamefont {Merkt}},\ }\href@noop {}
  {\bibfield  {journal} {\bibinfo  {journal} {Phys. Rev.~A}\ }\textbf {\bibinfo
  {volume} {77}},\ \bibinfo {pages} {042502} (\bibinfo {year}
  {2008})}\BibitemShut {NoStop}%
\bibitem [{\citenamefont {Jungen}(2011)}]{jungen11a}%
  \BibitemOpen
  \bibfield  {author} {\bibinfo {author} {\bibfnamefont {{\relax
  Ch}.}~\bibnamefont {Jungen}},\ }\bibfield  {title} {\enquote {\bibinfo
  {title} {{Elements of Quantum Defect Theory}},}\ }in\ \href@noop {} {\emph
  {\bibinfo {booktitle} {Handbook of High-resolution Spectroscopy}}},\
  Vol.~\bibinfo {volume} {1},\ \bibinfo {editor} {edited by\ \bibinfo {editor}
  {\bibfnamefont {M.}~\bibnamefont {Quack}}\ and\ \bibinfo {editor}
  {\bibfnamefont {F.}~\bibnamefont {Merkt}}}\ (\bibinfo  {publisher} {John
  Wiley \& Sons},\ \bibinfo {address} {Hoboken},\ \bibinfo {year} {2011})\ p.\
  \bibinfo {pages} {471}\BibitemShut {NoStop}%
\bibitem [{\citenamefont {Sprecher}, \citenamefont {\mbox{Ch.} Jungen},\ and\
  \citenamefont {Merkt}(2014)}]{sprecher14a}%
  \BibitemOpen
  \bibfield  {author} {\bibinfo {author} {\bibfnamefont {D.}~\bibnamefont
  {Sprecher}}, \bibinfo {author} {\bibnamefont {\mbox{Ch.} Jungen}}, \ and\
  \bibinfo {author} {\bibfnamefont {F.}~\bibnamefont {Merkt}},\ }\href@noop {}
  {\bibfield  {journal} {\bibinfo  {journal} {J. Chem. Phys.}\ }\textbf
  {\bibinfo {volume} {140}},\ \bibinfo {pages} {104303} (\bibinfo {year}
  {2014})}\BibitemShut {NoStop}%
\bibitem [{\citenamefont {Tong}, \citenamefont {Winney},\ and\ \citenamefont
  {Willitsch}(2010)}]{tong10a}%
  \BibitemOpen
  \bibfield  {author} {\bibinfo {author} {\bibfnamefont {X.}~\bibnamefont
  {Tong}}, \bibinfo {author} {\bibfnamefont {A.~H.}\ \bibnamefont {Winney}}, \
  and\ \bibinfo {author} {\bibfnamefont {S.}~\bibnamefont {Willitsch}},\
  }\href@noop {} {\bibfield  {journal} {\bibinfo  {journal} {Phys. Rev. Lett.}\
  }\textbf {\bibinfo {volume} {105}},\ \bibinfo {pages} {143001} (\bibinfo
  {year} {2010})}\BibitemShut {NoStop}%
\bibitem [{\citenamefont {Tong}\ \emph {et~al.}(2012)\citenamefont {Tong},
  \citenamefont {Nagy}, \citenamefont {\mbox{Yosa Reyes}}, \citenamefont
  {Germann}, \citenamefont {Meuwly},\ and\ \citenamefont
  {Willitsch}}]{tong12a}%
  \BibitemOpen
  \bibfield  {author} {\bibinfo {author} {\bibfnamefont {X.}~\bibnamefont
  {Tong}}, \bibinfo {author} {\bibfnamefont {T.}~\bibnamefont {Nagy}}, \bibinfo
  {author} {\bibfnamefont {J.}~\bibnamefont {\mbox{Yosa Reyes}}}, \bibinfo
  {author} {\bibfnamefont {M.}~\bibnamefont {Germann}}, \bibinfo {author}
  {\bibfnamefont {M.}~\bibnamefont {Meuwly}}, \ and\ \bibinfo {author}
  {\bibfnamefont {S.}~\bibnamefont {Willitsch}},\ }\href@noop {} {\bibfield
  {journal} {\bibinfo  {journal} {Chem. Phys. Lett.}\ }\textbf {\bibinfo
  {volume} {547}},\ \bibinfo {pages} {1} (\bibinfo {year} {2012})}\BibitemShut
  {NoStop}%
\bibitem [{\citenamefont {Germann}, \citenamefont {Tong},\ and\ \citenamefont
  {Willitsch}(2014)}]{germann14a}%
  \BibitemOpen
  \bibfield  {author} {\bibinfo {author} {\bibfnamefont {M.}~\bibnamefont
  {Germann}}, \bibinfo {author} {\bibfnamefont {X.}~\bibnamefont {Tong}}, \
  and\ \bibinfo {author} {\bibfnamefont {S.}~\bibnamefont {Willitsch}},\
  }\href@noop {} {\bibfield  {journal} {\bibinfo  {journal} {Nat. Phys.}\
  }\textbf {\bibinfo {volume} {10}},\ \bibinfo {pages} {820} (\bibinfo {year}
  {2014})}\BibitemShut {NoStop}%
\bibitem [{\citenamefont {Brown}\ and\ \citenamefont
  {Carrington}(2003)}]{brown03a}%
  \BibitemOpen
  \bibfield  {author} {\bibinfo {author} {\bibfnamefont {J.~M.}\ \bibnamefont
  {Brown}}\ and\ \bibinfo {author} {\bibfnamefont {A.}~\bibnamefont
  {Carrington}},\ }\href@noop {} {\emph {\bibinfo {title} {Rotational
  Spectroscopy of Diatomic Molecules}}}\ (\bibinfo  {publisher} {Cambridge
  University Press},\ \bibinfo {address} {Cambridge},\ \bibinfo {year}
  {2003})\BibitemShut {NoStop}%
\bibitem [{\citenamefont {Germann}\ and\ \citenamefont
  {Willitsch}()}]{germann16cPROVISORISCH_arXiv}%
  \BibitemOpen
  \bibfield  {author} {\bibinfo {author} {\bibfnamefont {M.}~\bibnamefont
  {Germann}}\ and\ \bibinfo {author} {\bibfnamefont {S.}~\bibnamefont
  {Willitsch}},\ }\href@noop {} {\bibinfo  {journal} {(to be published)}\
  }\BibitemShut {NoStop}%
\bibitem [{\citenamefont {Braunstein}, \citenamefont {McKoy},\ and\
  \citenamefont {Dixit}(1992)}]{braunstein92a}%
  \BibitemOpen
\bibfield  {journal} {  }\bibfield  {author} {\bibinfo {author} {\bibfnamefont
  {M.}~\bibnamefont {Braunstein}}, \bibinfo {author} {\bibfnamefont
  {V.}~\bibnamefont {McKoy}}, \ and\ \bibinfo {author} {\bibfnamefont {S.~N.}\
  \bibnamefont {Dixit}},\ }\href@noop {} {\bibfield  {journal} {\bibinfo
  {journal} {J.~Chem. Phys.}\ }\textbf {\bibinfo {volume} {96}},\ \bibinfo
  {pages} {5726} (\bibinfo {year} {1992})}\BibitemShut {NoStop}%
\bibitem [{\citenamefont {Zare}(1988)}]{zare88a}%
  \BibitemOpen
  \bibfield  {author} {\bibinfo {author} {\bibfnamefont {R.~N.}\ \bibnamefont
  {Zare}},\ }\href@noop {} {\emph {\bibinfo {title} {Angular Momentum}}}\
  (\bibinfo  {publisher} {John Wiley \& Sons},\ \bibinfo {address} {New York},\
  \bibinfo {year} {1988})\BibitemShut {NoStop}%
\bibitem [{\citenamefont {Varshalovich}, \citenamefont {Moskalev},\ and\
  \citenamefont {Khersonskii}(1989)}]{varshalovich89a}%
  \BibitemOpen
  \bibfield  {author} {\bibinfo {author} {\bibfnamefont {D.~A.}\ \bibnamefont
  {Varshalovich}}, \bibinfo {author} {\bibfnamefont {A.~N.}\ \bibnamefont
  {Moskalev}}, \ and\ \bibinfo {author} {\bibfnamefont {V.~K.}\ \bibnamefont
  {Khersonskii}},\ }\href@noop {} {\emph {\bibinfo {title} {Quantum Theory of
  Angular Momentum}}}\ (\bibinfo  {publisher} {World Scientific},\ \bibinfo
  {address} {Singapore},\ \bibinfo {year} {1989})\BibitemShut {NoStop}%
\bibitem [{\citenamefont {Edmonds}(1964)}]{edmonds64a}%
  \BibitemOpen
  \bibfield  {author} {\bibinfo {author} {\bibfnamefont {A.~R.}\ \bibnamefont
  {Edmonds}},\ }\href@noop {} {\emph {\bibinfo {title} {Drehimpulse in der
  Quantenmechanik}}}\ (\bibinfo  {publisher} {Bibliographisches Institut},\
  \bibinfo {address} {Mannheim},\ \bibinfo {year} {1964})\BibitemShut {NoStop}%
\bibitem [{\citenamefont {Germann}(2016)}]{germann16a}%
  \BibitemOpen
  \bibfield  {author} {\bibinfo {author} {\bibfnamefont {M.}~\bibnamefont
  {Germann}},\ }\href@noop {} {Ph.D. thesis},\ \bibinfo  {school} {University
  of Basel} (\bibinfo {year} {2016})\BibitemShut {NoStop}%
\bibitem [{\citenamefont {Dixit}\ \emph {et~al.}(1985)\citenamefont {Dixit},
  \citenamefont {Lynch}, \citenamefont {McKoy},\ and\ \citenamefont
  {Huo}}]{dixit85b}%
  \BibitemOpen
  \bibfield  {author} {\bibinfo {author} {\bibfnamefont {S.~N.}\ \bibnamefont
  {Dixit}}, \bibinfo {author} {\bibfnamefont {D.~L.}\ \bibnamefont {Lynch}},
  \bibinfo {author} {\bibfnamefont {V.}~\bibnamefont {McKoy}}, \ and\ \bibinfo
  {author} {\bibfnamefont {W.~M.}\ \bibnamefont {Huo}},\ }\href@noop {}
  {\bibfield  {journal} {\bibinfo  {journal} {Phys. Rev.~A}\ }\textbf {\bibinfo
  {volume} {32}},\ \bibinfo {pages} {1267} (\bibinfo {year}
  {1985})}\BibitemShut {NoStop}%
\bibitem [{\citenamefont {Frosch}\ and\ \citenamefont
  {Foley}(1952)}]{frosch52a}%
  \BibitemOpen
  \bibfield  {author} {\bibinfo {author} {\bibfnamefont {R.~A.}\ \bibnamefont
  {Frosch}}\ and\ \bibinfo {author} {\bibfnamefont {H.~M.}\ \bibnamefont
  {Foley}},\ }\href@noop {} {\bibfield  {journal} {\bibinfo  {journal} {Phys.
  Rev.}\ }\textbf {\bibinfo {volume} {88}},\ \bibinfo {pages} {1337} (\bibinfo
  {year} {1952})}\BibitemShut {NoStop}%
\bibitem [{\citenamefont {Dunn}(1972)}]{dunn72a}%
  \BibitemOpen
  \bibfield  {author} {\bibinfo {author} {\bibfnamefont {T.~M.}\ \bibnamefont
  {Dunn}},\ }\bibfield  {title} {\enquote {\bibinfo {title} {{Nuclear Hyperfine
  Structure in the Electronic Spectra of Diatomic Molecules}},}\ }in\
  \href@noop {} {\emph {\bibinfo {booktitle} {Molecular Spectroscopy: Modern
  Research}}},\ \bibinfo {editor} {edited by\ \bibinfo {editor} {\bibfnamefont
  {K.}~\bibnamefont {Narahari~Rao}}\ and\ \bibinfo {editor} {\bibfnamefont
  {C.}~\bibnamefont {Weldon~Mathews}}}\ (\bibinfo  {publisher} {Academic
  Press},\ \bibinfo {address} {New York},\ \bibinfo {year} {1972})\BibitemShut
  {NoStop}%
\bibitem [{\citenamefont {Hsu}\ \emph {et~al.}(1998)\citenamefont {Hsu},
  \citenamefont {Evans}, \citenamefont {Stimson}, \citenamefont {Ng},\ and\
  \citenamefont {Heimann}}]{hsu98a}%
  \BibitemOpen
  \bibfield  {author} {\bibinfo {author} {\bibfnamefont {C.-W.}\ \bibnamefont
  {Hsu}}, \bibinfo {author} {\bibfnamefont {M.}~\bibnamefont {Evans}}, \bibinfo
  {author} {\bibfnamefont {S.}~\bibnamefont {Stimson}}, \bibinfo {author}
  {\bibfnamefont {C.~Y.}\ \bibnamefont {Ng}}, \ and\ \bibinfo {author}
  {\bibfnamefont {P.}~\bibnamefont {Heimann}},\ }\href@noop {} {\bibfield
  {journal} {\bibinfo  {journal} {Chem. Phys.}\ }\textbf {\bibinfo {volume}
  {231}},\ \bibinfo {pages} {121} (\bibinfo {year} {1998})}\BibitemShut
  {NoStop}%
\bibitem [{\citenamefont {Rosner}, \citenamefont {Gaily},\ and\ \citenamefont
  {Holt}(1985)}]{rosner85a}%
  \BibitemOpen
  \bibfield  {author} {\bibinfo {author} {\bibfnamefont {S.~D.}\ \bibnamefont
  {Rosner}}, \bibinfo {author} {\bibfnamefont {T.~D.}\ \bibnamefont {Gaily}}, \
  and\ \bibinfo {author} {\bibfnamefont {R.~A.}\ \bibnamefont {Holt}},\
  }\href@noop {} {\bibfield  {journal} {\bibinfo  {journal} {J. Mol.
  Spectrosc.}\ }\textbf {\bibinfo {volume} {109}},\ \bibinfo {pages} {73}
  (\bibinfo {year} {1985})}\BibitemShut {NoStop}%
\bibitem [{\citenamefont {Freund}\ \emph {et~al.}(1970)\citenamefont {Freund},
  \citenamefont {Miller}, \citenamefont {{De Santis}},\ and\ \citenamefont
  {Lurio}}]{freund70a}%
  \BibitemOpen
  \bibfield  {author} {\bibinfo {author} {\bibfnamefont {R.~S.}\ \bibnamefont
  {Freund}}, \bibinfo {author} {\bibfnamefont {T.~A.}\ \bibnamefont {Miller}},
  \bibinfo {author} {\bibfnamefont {D.}~\bibnamefont {{De Santis}}}, \ and\
  \bibinfo {author} {\bibfnamefont {A.}~\bibnamefont {Lurio}},\ }\href@noop {}
  {\bibfield  {journal} {\bibinfo  {journal} {J. Chem. Phys.}\ }\textbf
  {\bibinfo {volume} {53}},\ \bibinfo {pages} {2290} (\bibinfo {year}
  {1970})}\BibitemShut {NoStop}%
\bibitem [{\citenamefont {{De Santis}}\ \emph {et~al.}(1973)\citenamefont {{De
  Santis}}, \citenamefont {Lurio}, \citenamefont {Miller},\ and\ \citenamefont
  {Freund}}]{desantis73a}%
  \BibitemOpen
  \bibfield  {author} {\bibinfo {author} {\bibfnamefont {D.}~\bibnamefont {{De
  Santis}}}, \bibinfo {author} {\bibfnamefont {A.}~\bibnamefont {Lurio}},
  \bibinfo {author} {\bibfnamefont {T.~A.}\ \bibnamefont {Miller}}, \ and\
  \bibinfo {author} {\bibfnamefont {R.~S.}\ \bibnamefont {Freund}},\
  }\href@noop {} {\bibfield  {journal} {\bibinfo  {journal} {J. Chem. Phys.}\
  }\textbf {\bibinfo {volume} {58}},\ \bibinfo {pages} {4625} (\bibinfo {year}
  {1973})}\BibitemShut {NoStop}%
\bibitem [{\citenamefont {Legon}\ and\ \citenamefont
  {Fowler}(1992)}]{legon92a}%
  \BibitemOpen
  \bibfield  {author} {\bibinfo {author} {\bibfnamefont {A.~C.}\ \bibnamefont
  {Legon}}\ and\ \bibinfo {author} {\bibfnamefont {P.~W.}\ \bibnamefont
  {Fowler}},\ }\href@noop {} {\bibfield  {journal} {\bibinfo  {journal} {Z.
  Naturforsch.}\ }\textbf {\bibinfo {volume} {47a}},\ \bibinfo {pages} {367}
  (\bibinfo {year} {1992})}\BibitemShut {NoStop}%
\bibitem [{\citenamefont {Baltzer}, \citenamefont {Karlsson},\ and\
  \citenamefont {Wannberg}(1992)}]{baltzer92a}%
  \BibitemOpen
  \bibfield  {author} {\bibinfo {author} {\bibfnamefont {P.}~\bibnamefont
  {Baltzer}}, \bibinfo {author} {\bibfnamefont {L.}~\bibnamefont {Karlsson}}, \
  and\ \bibinfo {author} {\bibfnamefont {B.}~\bibnamefont {Wannberg}},\
  }\href@noop {} {\bibfield  {journal} {\bibinfo  {journal} {Phys. Rev.~A}\
  }\textbf {\bibinfo {volume} {46}},\ \bibinfo {pages} {315} (\bibinfo {year}
  {1992})}\BibitemShut {NoStop}%
\bibitem [{\citenamefont {\"Ohrwall}, \citenamefont {Baltzer},\ and\
  \citenamefont {Bozek}(1999)}]{oehrwall99a}%
  \BibitemOpen
  \bibfield  {author} {\bibinfo {author} {\bibfnamefont {G.}~\bibnamefont
  {\"Ohrwall}}, \bibinfo {author} {\bibfnamefont {P.}~\bibnamefont {Baltzer}},
  \ and\ \bibinfo {author} {\bibfnamefont {J.}~\bibnamefont {Bozek}},\
  }\href@noop {} {\bibfield  {journal} {\bibinfo  {journal} {Phys. Rev. A}\
  }\textbf {\bibinfo {volume} {59}},\ \bibinfo {pages} {1903} (\bibinfo {year}
  {1999})}\BibitemShut {NoStop}%
\bibitem [{\citenamefont {Germann}\ and\ \citenamefont
  {Willitsch}(2016)}]{germann16b}%
  \BibitemOpen
  \bibfield  {author} {\bibinfo {author} {\bibfnamefont {M.}~\bibnamefont
  {Germann}}\ and\ \bibinfo {author} {\bibfnamefont {S.}~\bibnamefont
  {Willitsch}},\ }\href@noop {} {\bibfield  {journal} {\bibinfo  {journal}
  {Mol. Phys.}\ }\textbf {\bibinfo {volume} {114}},\ \bibinfo {pages} {769}
  (\bibinfo {year} {2016})}\BibitemShut {NoStop}%
\bibitem [{\citenamefont {Merkt}\ and\ \citenamefont
  {Softley}(1993)}]{merkt93c}%
  \BibitemOpen
  \bibfield  {author} {\bibinfo {author} {\bibfnamefont {F.}~\bibnamefont
  {Merkt}}\ and\ \bibinfo {author} {\bibfnamefont {T.~P.}\ \bibnamefont
  {Softley}},\ }\href@noop {} {\bibfield  {journal} {\bibinfo  {journal} {Int.
  Rev. Phys. Chem.}\ }\textbf {\bibinfo {volume} {12}},\ \bibinfo {pages} {205}
  (\bibinfo {year} {1993})}\BibitemShut {NoStop}%
\end{thebibliography}%

\end{document}